\documentclass[twocolumn]{aastex631}  
\usepackage{enumitem}
\usepackage{xcolor}
\usepackage{tablefootnote}
\usepackage{bm}
\usepackage{lineno}
\hypersetup{backref=true,pagebackref=true,  hyperindex=true,colorlinks=blue,breaklinks=true,  urlcolor=violet,linkcolor= red, bookmarks=true, 	bookmarksopen=true,  filecolor=cyan, citecolor=blue, linkbordercolor=blue}
\newcommand\wfirst{{Roman}}
\shorttitle{Parallax in microlensing due to free-floating planets}
\shortauthors{}

\begin{document}
\title{Parallax Effect in Microlensing Events due to Free-Floating Planets}

\author[0009-0007-4190-1269]{Parisa Sangtarash}
\affiliation{Department of Physics, Isfahan University of Technology, Isfahan 84156-83111, Iran}

\author[0000-0002-0167-3595]{Sedighe~Sajadian$^{\star}$}
\affiliation{Department of Physics, Isfahan University of Technology, Isfahan 84156-83111, Iran}

\footnote{$^{\star}$\textcolor{blue}{$\rm{e}$-$\rm{mail:s.sajadian@iut.ac.ir}$}}

\begin{abstract}
One of most important applications of microlensing observations is detecting free-floating planets(FFPs). The time scale of microlensing due to FFPs ($t_{\rm E}$) is short (a few days). Discerning the annual parallax effect in observations from these short-duration events by one observer is barely possible, though their parallax amplitude is larger than that in common events. In microlensing events due to FFPs, the lens-source relative trajectory alters because of the observer's motion by $\boldsymbol{\delta u}$. This deviation is a straight line if $t_{\rm E} \ll P_{\oplus}$, and its size is $\delta u\propto \pi_{\rm{rel}}$ ($P_{\oplus}$ is the observer's orbital period). So, most of observed microlensing events due to close FFPs have simple Paczy\'nsky lightcurves with indiscernible and valuable parallax. To evaluate destructive effects of invisible parallax in such events, we simulate $\sim9650$ microlensing events due to FFPs with $t_{\rm E}<10$ days that are observed only by The Nancy Grace Roman Space Telescope(\wfirst). We conclude that in half of these microlensing events the missing parallax alters the real lightcurves, changing their shape and derived properties(by $\Delta \chi^{2}\gtrsim100$). By fitting Paczy\'nski lightcurves to these affected events we evaluate the relative and dimensionless deviations in the lensing parameters from their real values ($\delta t_{\rm E}, \delta \rho_{\star}, ...$). We conclude that around $46$ FFPs which are discovered by \wfirst\ have lightcurves highly affected by invisible parallax with $\delta t_{\rm E}>0.1~\rm{and}~\delta \rho_{\star}>0.1$. Our study reveals the importance of simultaneous and dense observations of microlensing events viewed by \wfirst\ by other observers rotating the Sun in different orbits.
\end{abstract}


\section{Introduction}
The annual parallax effect is refereed to the Earth (the observer) rotation around the Sun which makes additional and apparent motions for close stars on the sky plane. Due to the parallax effect, the angular position of a star at the distance one kpc from us will alter by $2$ mas during $6$ months as measured from the Earth. Based on these apparent displacements of close stars in the sky plane, people even determine their distance from the Earth. Therefore, in astrophysical text books, the parallax effect has been introduced as one of primary methods for measuring distances of close stars \citep[e.g., ][]{2006bookCarroll}. In this way, the Hipparcos and Gaia telescopes have measured the distances of over a hundred thousand stars, and around $1.5$ billion stars in the Galaxy based on the parallax measurements \citep{1997Hipparcos,2023GDR3}.      

Additionally, the parallax effect alters the transient astrophysical events from the Galactic stars, because it alters stellar trajectories from straight to cycloid ones \citep[see, e.g.,][]{1995ApJAlcock}. The parallax-induced deviations in such events depend on two factors: (a) the stellar distances from the observer, and (b) durations of events. The angular extra motions of stars due to the parallax effect are scaled by $\pi_{\star}=\rm{au}/D$, where $D$ is the stellar distance from the observer. Also, the maximum displacements owing to the parallax effect occur during $6$ months. Hence, in short-duration events (e.g., 1-2 days) from far stars the parallax-induced deviations and bending of stellar trajectories from straight ones are barely recognizable.

One example of transient astrophysical phenomena is the gravitational microlensing. It refers to the temporary brightness increase of a background star in the Galaxy, because its light is passing from the gravitational potential of a foreground and collinear object \citep{Einstein1936,1964MNRASrefsdal,Liebes1964}. The most-common lenses in the Gravitational microlensing events toward the Galactic bulge are M-dwarfs with the masses $\sim 0.3~M_{\odot}$. These lens objects make microlensing events which last on average $\sim 20$-$30$ days \citep[see, e.g., ][]{gaudi2012,2008Dominik}. In a simple microlensing event the magnification factor depends on three parameters which are: (i) $\rho_{\star}$ the source radius normalized to the Einstein radius and projected on the sky plane, (ii) the brightness profile of the source disk, and (iii) $u$ the lens-source relative trajectory normalized to the Einstein radius. Here, the Einstein radius is the radius of the images' ring at the time of complete alignment between the lens, source and the observer lines of sight. 

\noindent In these events, the parallax effect alters $u$ from a straight line to a cycloid one. The amplitude of parallax-induced deviations in the lens-source trajectory is $\pi_{\rm E}$ (the so-called parallax amplitude). Discerning this parallax amplitude helps to resolve the known microlensing degeneracy, since $\pi_{\rm E}$ is a degenerate function of the lens mass and its distance \citep[e.g., ][]{1966Refesdal,1992Gould,2005Smith}. In joint ground-based and space-based observations from microlensing events, measuring up to three physical parameters of lenses is possible because of different parallax effects from different observers \citep[][]{1994ApJGould,1995ApJGould,1999ApJGould}. As some examples of realizing satellite parallax in real microlensing observations one can study \citet{2002ApJAn,2019AJYossi,2020ApJZang,2020AJHirao}.  

One special class of microlensing events are the ones due to free-floating planets (FFPs) which have short durations (up to a few days) \citep[e.g., ][]{Sumi2011Natur, Mroz2017Natur, 2020AJMroz,2020Mroz_1,2023AJkoshimoto}. In these events (i) the parallax amplitude is larger than that in common microlensing events due to M-dwarfs, and (ii) the observer trajectory with respect to the Sun is almost a straight line with ignorable bending. Therefore, in such events the straight lens-source relative trajectory varies with a considerable and straight deviation, i.e., $\boldsymbol{u}_{\odot}+\boldsymbol{\delta u}$. Accordingly, the parallax effect in these events alters the real light curves to other simple ones with indistinguishable parallax effects.

A considerable number of short-duration microlensing events due to FFPs will be discovered by the \wfirst\ Space Telescope during its microlensing survey \citep[see, e.g., ][]{2019ApJPenny,2020AJjohnson}. We note that \citet{2023AJ108sumi} presented the first measurement of the mass function of free-floating planets and predicted that \wfirst\ could detect a significant number of these objects with masses down to that of Mars. Since measuring their parallax effects is crucial to correctly extract the lens mass and its distance, possible joint observations by \wfirst,\ LSST \citep{2009LSSTbook}, Euclid, and Chinese Space Station telescopes from these events were proposed by \citet{2019ApJBachelt,2020Ban,2022AABachelt,2022RAAZhu}. Also, \citet{2013Hamolli} simulated microlensing events due to FFPs by considering the parallax effect and discussed on the best position of the Earth in its orbit around the Sun for measuring the parallax effect. 

Simultaneous observations are not possible for some of short-duration microlensing events. For instance, for the ones that either (i) their durations are very short, or (ii) their observing field is not observable by two telescopes simultaneously, or (iii) other telescopes have other priorities for their observations. Modeling such microlensing events leads to lensing parameters somewhat deviated from their real amounts. We study this point (how much invisible parallax is destructive while interpreting short-duration events due to FFPs) in this work numerically, and statistically evaluate the parallax-induced deviations on the inferred lensing parameters from observations.   

In Section \ref{sec1}, we review the formalism for microlensing light curves by considering the parallax effect. In Section \ref{sec2}, we generate short-duration ($t_{\rm E}<10$ days) microlensing events due to FFPs toward the Galactic bulge, and assume these events are observed only by the \wfirst\ telescope during one $62$-day observing season. We extract the parallax-affected events by evaluating $\Delta \chi^{2}$ values. To calculate the parallax-induced deviations in their lensing parameters, we fit simple Paczy\'nski microlensing models to their simulated synthetic data. The statistical results are explained in Section \ref{sec3}. In the last section, we explain the results and conclusions.         

\section{Formalism: Microlensing and Parallax}\label{sec1}
In a microlensing event due to a point-like source star, the magnification factor only depends on the lens-source relative distance projected on the sky plane and normalized to the Einstein radius, $u$, as given by:  
\begin{eqnarray}
A(u)= \frac{u^{2}+2}{u\sqrt{u^{2}+4}}. 
\end{eqnarray}    

We first assume the observer is on the Sun's center (the so-called heliocentric frame). The heliocentric coordinate system is specified by three normal axes, i.e., $(\hat{n}_{1},~\hat{n}_{2},~\hat{z})$, where $\hat{z}$ is in the line of sight direction and toward the source star. $(\hat{n}_{1},~\hat{n}_{2})$ describe the sky plane, where $\hat{n}_{1}$ is in the direction of increasing the Galactic longitude, and $\hat{n}_{2}$ is toward the Galactic north pole. In this system, the components of lens-source relative trajectory projected on the sky plane $\boldsymbol{u}_{\odot}$ are given by: 
\begin{eqnarray}
u_{\odot,\hat{n}_{1}}&=& \frac{t-t_{0}}{t_{\rm E}} \cos \xi-u_{0}~\sin \xi, \nonumber \\
u_{\odot,\hat{n}_{2}}&=& \frac{t-t_{0}}{t_{\rm E}} \sin \xi+u_{0}~\cos \xi,
\end{eqnarray} 
where, $u_{\odot}=\Big(u_{0}^{2}+ (t-t_{0})^{2}/t_{\rm{E}}^{2} \Big)^{1/2}$, $t_{0}$ is the time of the closest approach, $u_{0}$ is the lens impact parameter, $\xi$ is the angle for projection of the lens-source relative trajectory on the $\hat{n}_{1}$ axis. $t_{\rm E}$ is the time to cross the angular Einstein radius with the lens-source relative velocity:
\begin{eqnarray}
t_{\rm E}=\frac{\theta_{\rm E}}{\mu_{\rm{rel}, \odot}},~~~~~\theta_{\rm E}=\frac{R_{\rm E}}{D_{\rm l}}=\sqrt{\frac{4~G~M_{\rm l}}{c^{2}}\frac{D_{\rm{ls}}}{D_{\rm l} D_{\rm s}}}
\end{eqnarray}
where, $G$ is the Gravitational constant, $c$ is the speed of light, $D_{\rm l}$, and $D_{\rm s}$ are the lens and source distances from the observer, and $D_{\rm{ls}}=D_{\rm s}-D_{\rm l}$. $\mu_{\rm{rel}, \odot}$ is the size of the lens-source angular velocity vector as measured in the heliocentric reference frame, which is given by:  
\begin{eqnarray}
\boldsymbol{\mu}_{\rm{rel}, \odot}= \frac{ \boldsymbol{v}_{\rm s, p}-\boldsymbol{v}_{\odot,p}}{D_{\rm s}}- \frac{\boldsymbol{v}_{\rm l, p}- \boldsymbol{v}_{\odot, p}}{D_{\rm l}},  
\end{eqnarray}
where $\boldsymbol{v}_{\rm l, p}$, $ \boldsymbol{v}_{\rm s, p}$, and $\boldsymbol{v}_{\odot, p}$ are the lens, source and the Sun velocity vectors projected on the sky plane. Since the Sun motion around the Galactic center during a lensing event is linear, the lens trajectory with respect to the source star (as measured in the heliocentric coordinate frame) is a straight line with a constant relative velocity $\boldsymbol{\mu}_{\rm{rel}, \odot}$. 

Now, we assume the observer is rotating the Sun. The annual motion of the observer around the Sun changes $\boldsymbol{\mu}_{\rm{rel}, \odot}$ by: 
\begin{eqnarray}
\boldsymbol{\mu}_{\rm{rel},~\rm{o}}=\boldsymbol{\mu}_{\rm{rel},~\odot} +\frac{\pi_{\rm{rel}}}{\rm{au}} \boldsymbol{v}_{\rm{o}, p}, 
\end{eqnarray}
where $\boldsymbol{v}_{\rm{o}, p}$ is the observer velocity with respect to the Sun projected on the sky plane, and $\pi_{\rm{rel}}=\rm{au}/D_{\rm l}-\rm{au}/D_{\rm s}$ is the relative parallax amplitude. Accordingly, the observer motion around the Sun changes two components of $\boldsymbol{u}_{\rm{o}}$ as:  
\begin{eqnarray}
u_{\rm{o}, \hat{n}_{1}}&=& u_{\odot, \hat{n}_{1}}+\pi_{\rm E} \int_{t_{0}}^{t} \frac{v_{\rm{o}, \hat{n}_{1}}}{\rm{au}} dt, \nonumber \\
u_{\rm{o}, \hat{n}_{2}}&=& u_{\odot, \hat{n}_{2}}+\pi_{\rm E}\int_{t_{0}}^{t} \frac{v_{\rm{o}, \hat{n}_{2}}}{\rm{au}}dt.
\label{upar}
\end{eqnarray}
These parallax-induced deviations alter periodically (same as a sine function) with the known amplitude $\pi_{\rm E}=\pi_{\rm{rel}}/\theta_{\rm E}=\sqrt{\pi_{\rm{rel}} \big/\kappa~M_{\rm l}}$, where $\kappa$ is a constant.

\begin{figure*}
\centering
\includegraphics[width=0.49\textwidth]{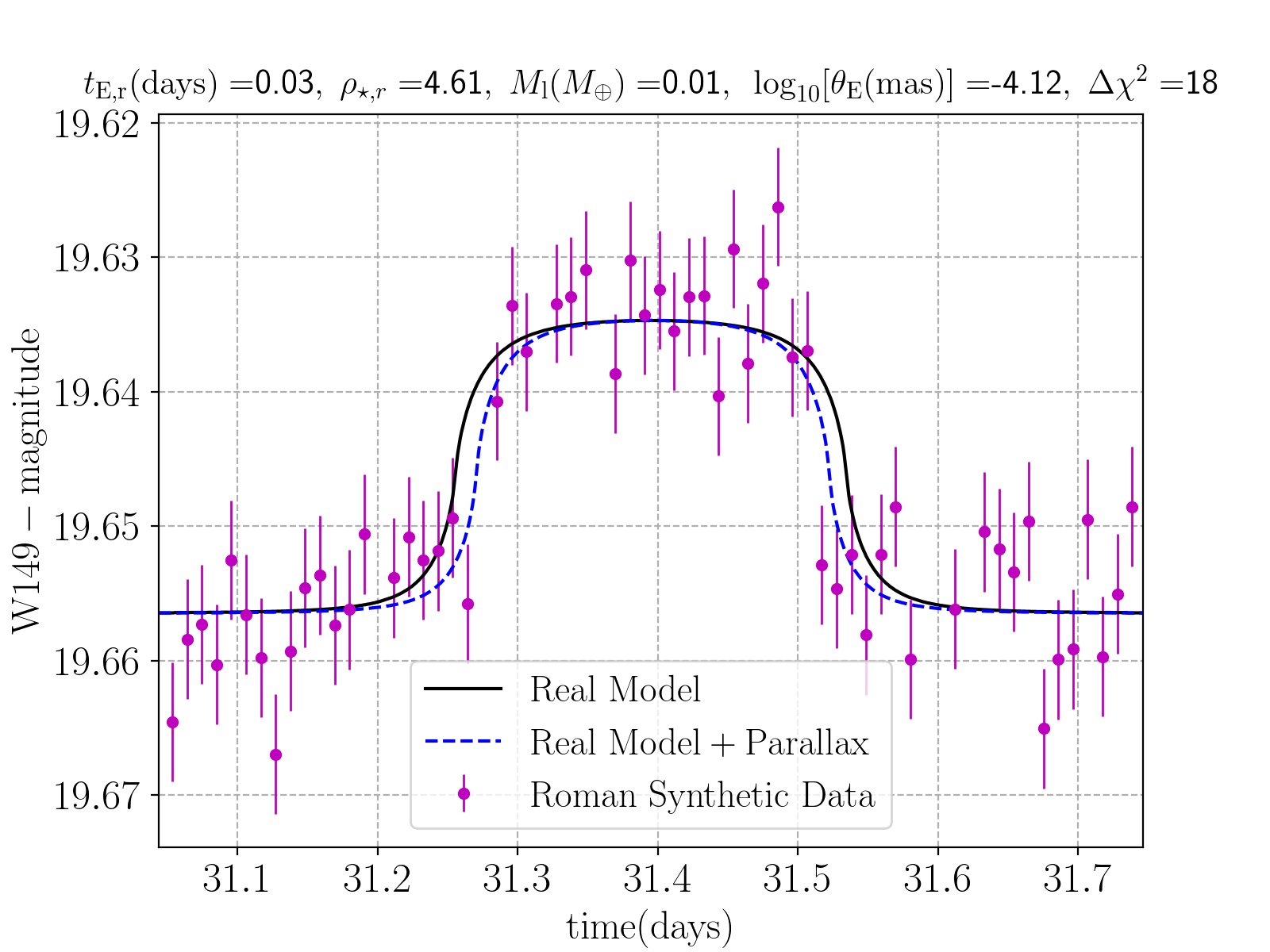}
\includegraphics[width=0.49\textwidth]{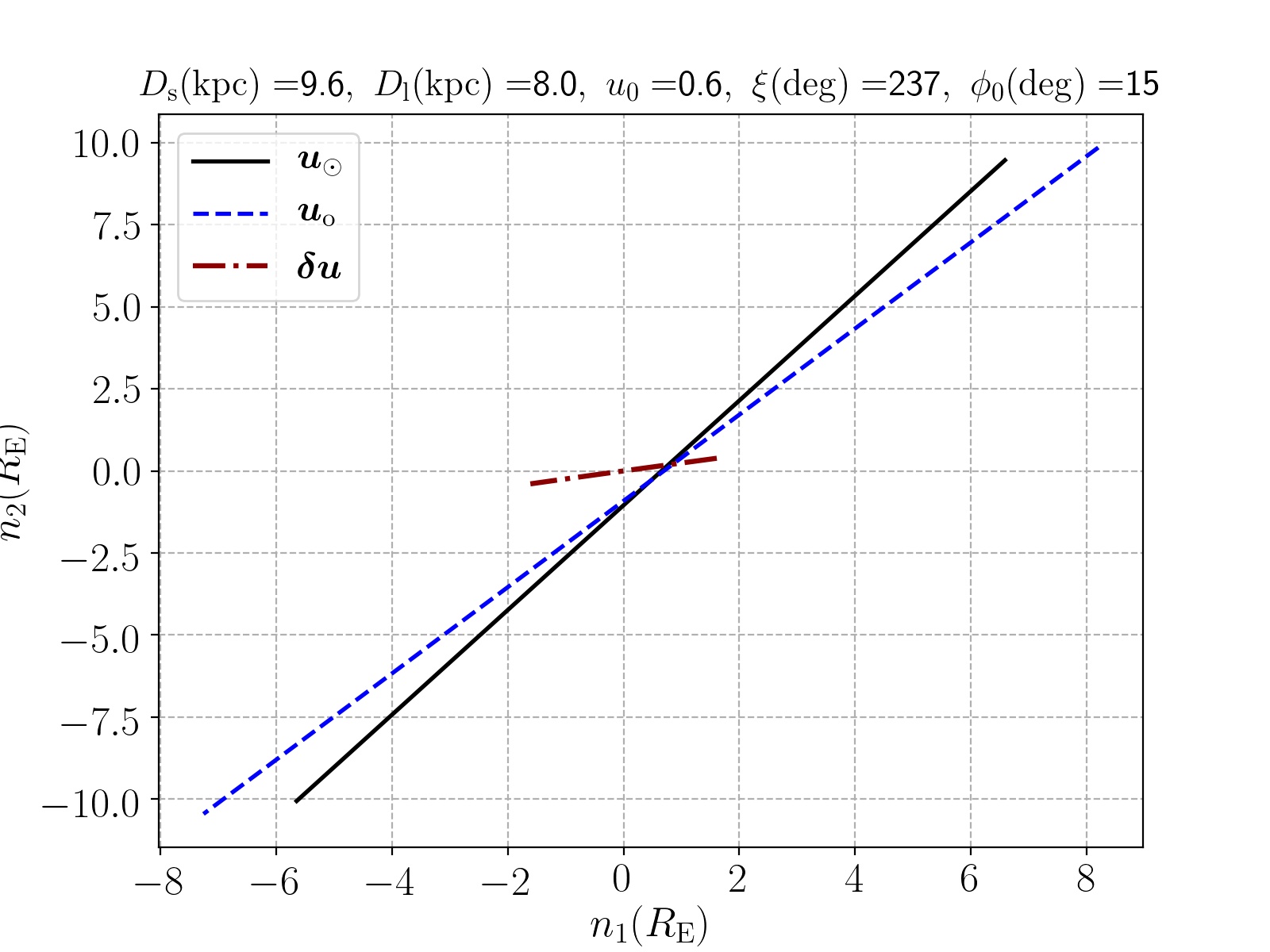}
\includegraphics[width=0.49\textwidth]{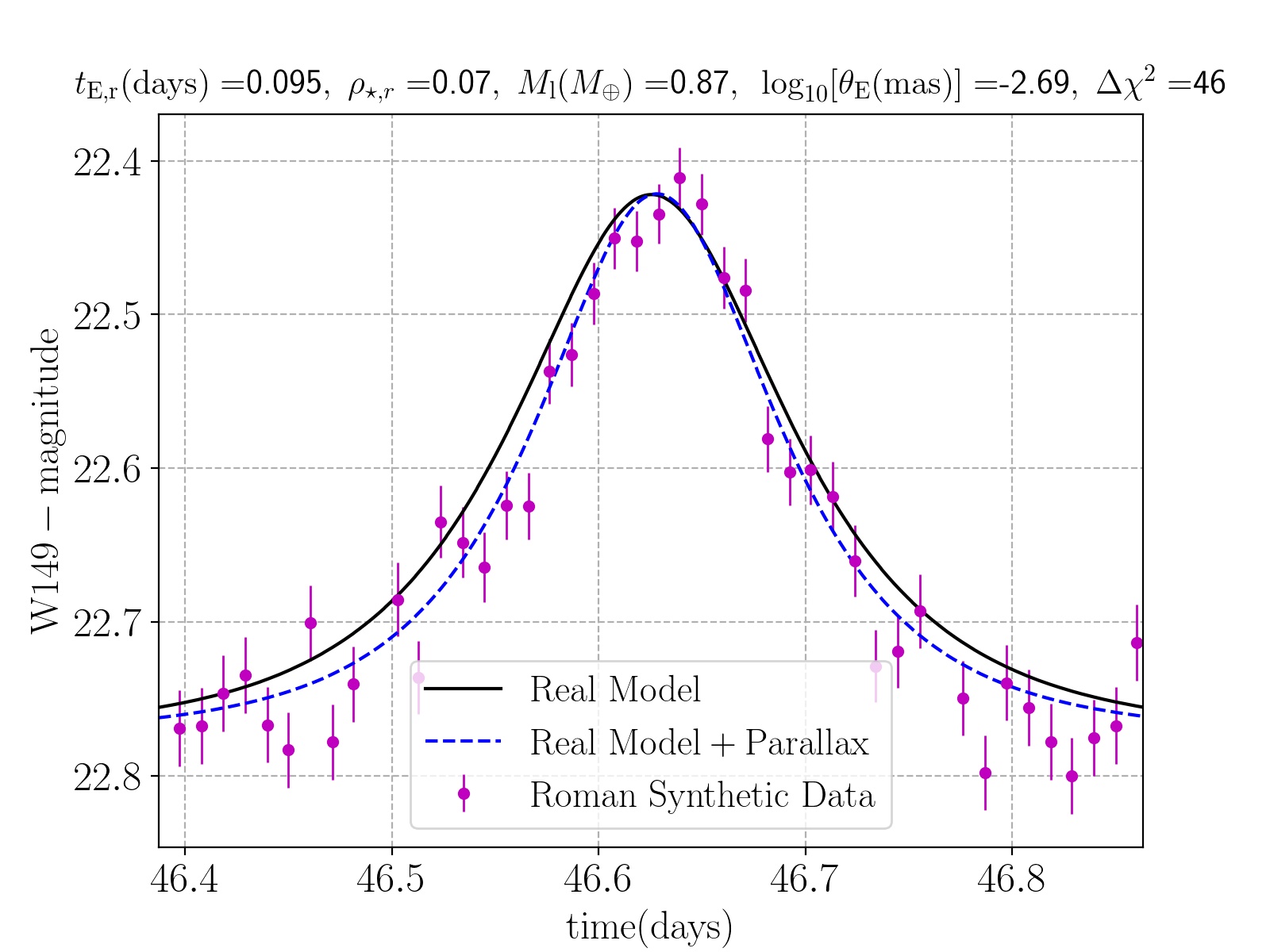}
\includegraphics[width=0.49\textwidth]{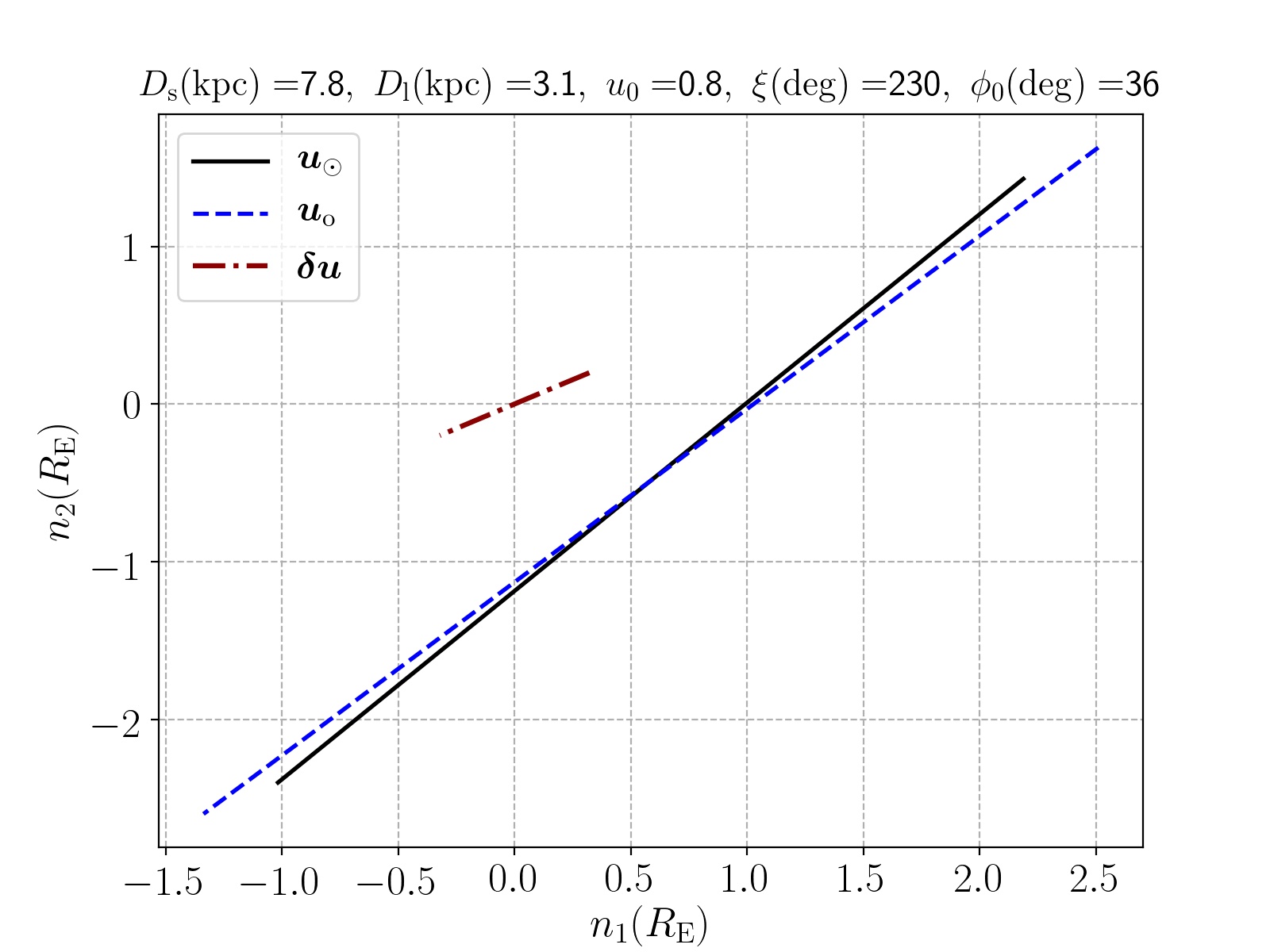}
\caption{Left panels show two examples of microlensing light curves due to FFPs in which the parallax effects are not discernible. Right panels contain their lens-source relative trajectories projected on the sky plane without (black solid lines $\boldsymbol{u}_{\odot}$) and with (blue dashed lines $\boldsymbol{u}_{\rm o}$) considering the parallax effect. In these panels, the dark-red dot-dashed lines are $\boldsymbol{\delta u}$. The index $r$ for parameters refers to their values in the real models. }\label{fig1}
\end{figure*}

\subsection{Case study: Short-duration Microlensing due to FFPs}
Here, we aim to evaluate the second terms in Equations \ref{upar} (the parallax-induced deviations) in short-duration microlensing events due to FFPs. To project the observer velocity vector on the sky, we first specify its components in the Galactic coordinate system. The Galactic coordinate system is described with three normal directions $(U,~V,~W)$, which are radial toward the Galactic center, clock-ward tangential in the Galactic plane, and normal to the Galactic plane and toward the Galactic north pole, respectively. The observer velocity components in this coordinate system is given by:   
\begin{eqnarray}
v_{\rm{o}, U}&=& +V_{\rm{o}} ~\sin \Omega~\cos \theta_{\oplus}, \nonumber \\
v_{\rm{o}, V}&=& -V_{\rm{o}} ~\cos \Omega, \nonumber\\
v_{\rm{o}, W}&=&-V_{\rm{o}} ~\sin \Omega~\sin \theta_{\oplus},
\label{uvw}
\end{eqnarray}
where, $\theta_{\oplus}=\pi/3$ is the angle between the Earth orbital plane around the Sun and the Galactic plane, $V_{\rm{o}}=R_{\rm{o}}~\omega$, $\Omega=\omega (t-t_{0}) +\phi_{0}$, $\omega=2\pi/P_{\oplus}$ is the Earth angular velocity around the Sun, and $P_{\oplus}=365.25$ days. $\phi_{0}$ is an initial phase which describes the Earth velocity's vector at the time of the closest approach in the lensing formalism. $R_{\rm o}$ is the orbital radius of the observer around the Sun, which is the astronomical unit (au) when the observer is the Earth and $1.01$ au for the \wfirst\ telescope (from L2 Lagrangian point). Here, we assumed that the observer is rotating the Sun in the Earth orbital plane and in a circular orbit. 

According to the formalism introduced in Appendix of \citet{2023sajadiansahu}, we derive the projected components of the Earth velocity on the sky plane (in the directions $\hat{n}_{1}, \hat{n}_{2}$) as: 
\begin{eqnarray}
v_{\rm{o}, \hat{n}_{1}}&=&\sin \alpha~v_{\rm{o}, U} - \cos \alpha~v_{\rm{o}, V}, \nonumber \\
v_{\rm{o}, \hat{n}_{2}}&=&\sin b \Big[\cos \alpha~v_{\rm{o}, U} + \sin\alpha~v_{\rm{o}, V} \Big] + \cos b~v_{\rm{o}, W}.
\label{voo}
\end{eqnarray}
Here, $b$ is the Galactic latitude of the line of sight of the source star,  $\alpha\simeq \pi-l$, and $l$ is the Galactic longitude of this line of sight. Toward the Galactic bulge, $l\simeq0$, $b\simeq 0$, and so $\alpha\simeq \pi$. Additionally, for short-duration microlensing events due to FFPs, $\big| t-t_{0}\big |\ll P_{\oplus}$. Considering these features and using Equations \ref{voo}, two parallax-induced deviations (given in Equations \ref{upar}) are:
\begin{eqnarray}
\delta u_{\hat{n}_{1}}&\simeq&-R'_{\rm o}\pi_{\rm E}\big( \sin\Omega-\sin \phi_{0}\big)\nonumber\\&\simeq& -R'_{\rm o}\pi_{\rm E}\big[\omega(t-t_{0}) \cos \phi_{0}-\frac{1}{2}\sin \phi_{0}\omega^{2}(t-t_{0})^2\big],\nonumber\\
\delta u_{\hat{n}_{2}}&\simeq& \frac{\sqrt{3}}{2}R'_{\rm o}\pi_{\rm E}\big( \cos\Omega -cos \phi_{0} \big)\nonumber\\&\simeq&\frac{\sqrt{3}}{2}R'_{\rm o}\pi_{\rm E}\big[-\omega(t-t_{0})\sin\phi_{0} -\frac{1}{2}\omega^{2}(t-t_{0})^{2}\cos\phi_{0}\big],
\label{twot}
\end{eqnarray}
where $R'_{\rm o}=R_{\rm o}/\rm{au}$, and $\boldsymbol{u}_{\rm{o}}=\boldsymbol{u}_{\odot}+\boldsymbol{\delta u}$. In these relations, although $\omega (t-t_{0})\sim t_{\rm E}/P_{\oplus}$ is very small, in short-duration microlensing events due to FFPs the parallax amplitudes ($\pi_{\rm E}\propto 1/\sqrt{M_{\rm l}}$)  are considerable and larger than those in the common microlensing events by $5$ up to $3000$ times for $M_{\rm l} \in [0.01~M_{\oplus}, 15~M_{\rm J}]$. 

Hence, the parallax effect in microlensing events due to FFPs is not discernible, because two parallax-induced deviation terms (Equation \ref{twot}) as far as $t_{\rm E}\ll P_{\oplus}$ behave linearly versus time without any bending and acceleration (by considering only first-order terms). In that case, $\boldsymbol{u}_{\rm o}$ will be a straight line and the offered light curve will be symmetric with respect to its peak. However, if $\pi_{\rm{E}}\times t_{\rm E}$ has the same order of magnitude with $P_{\oplus}$, i.e., $\pi_{\rm{E}}\times t_{\rm E} (=\pi_{\rm{rel}}/\mu_{\rm{rel}, \rm o}) \sim P_{\oplus}$, then $\boldsymbol{\delta u}$ alters $\boldsymbol{u}_{\odot}$ considerably, which means $A(u_{\rm o})$ and $A(u_{\odot})$ are different whereas both of them are simple and symmetric. In such events parallax effect is indistinguishable and invisible, whereas it changes the observed light curve and its parameters considerably. Accordingly, in short-duration events due to FFPs, $\delta u\propto \pi_{\rm{rel}}$ which means parallax can make large deviations when FFPs are close to the observer with high $\pi_{\rm{rel}}$ values. We examine these points in the next section based on a Monte Carlo simulation.  

\begin{figure*}
\includegraphics[width=0.33\textwidth]{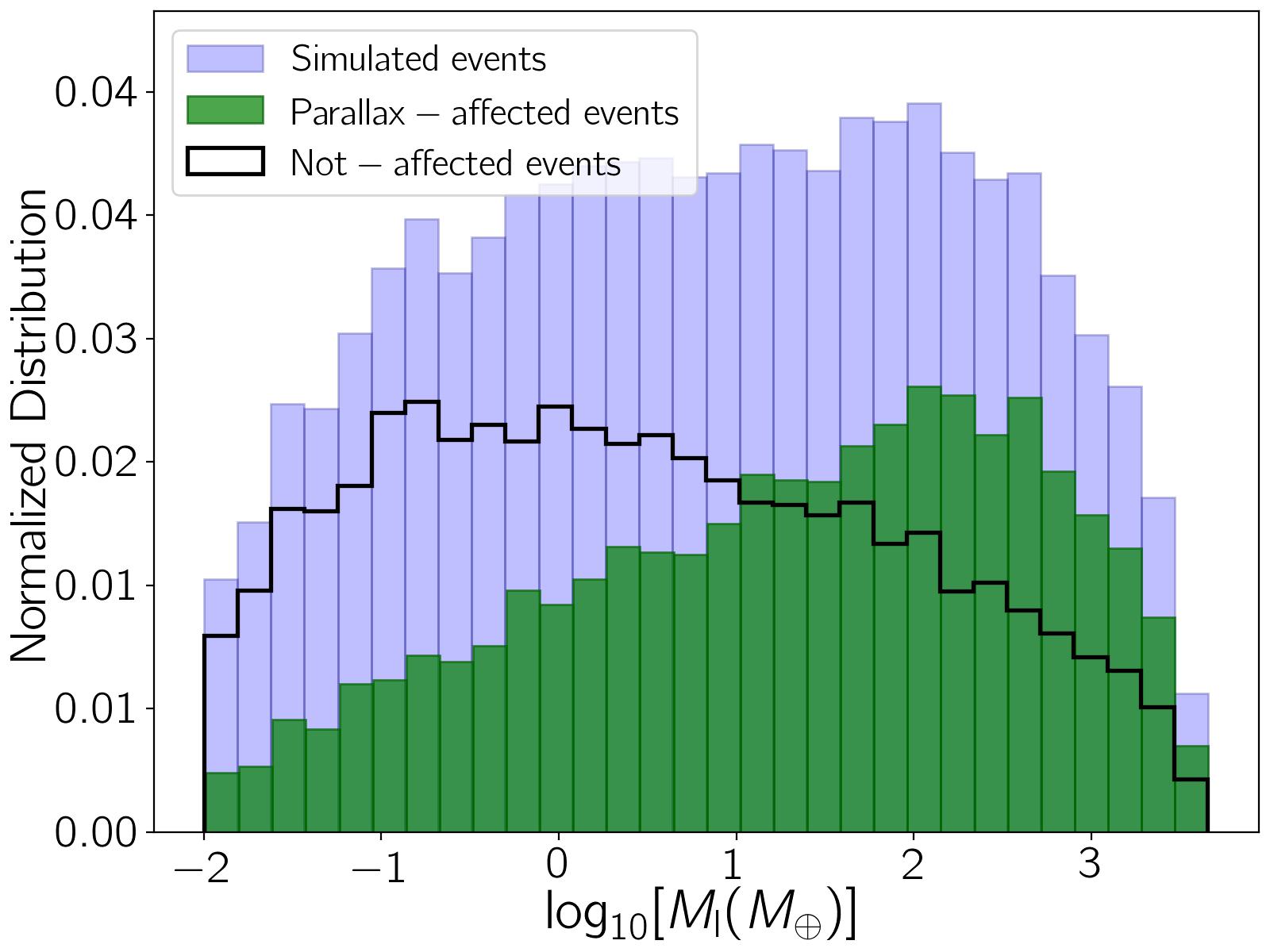}
\includegraphics[width=0.33\textwidth]{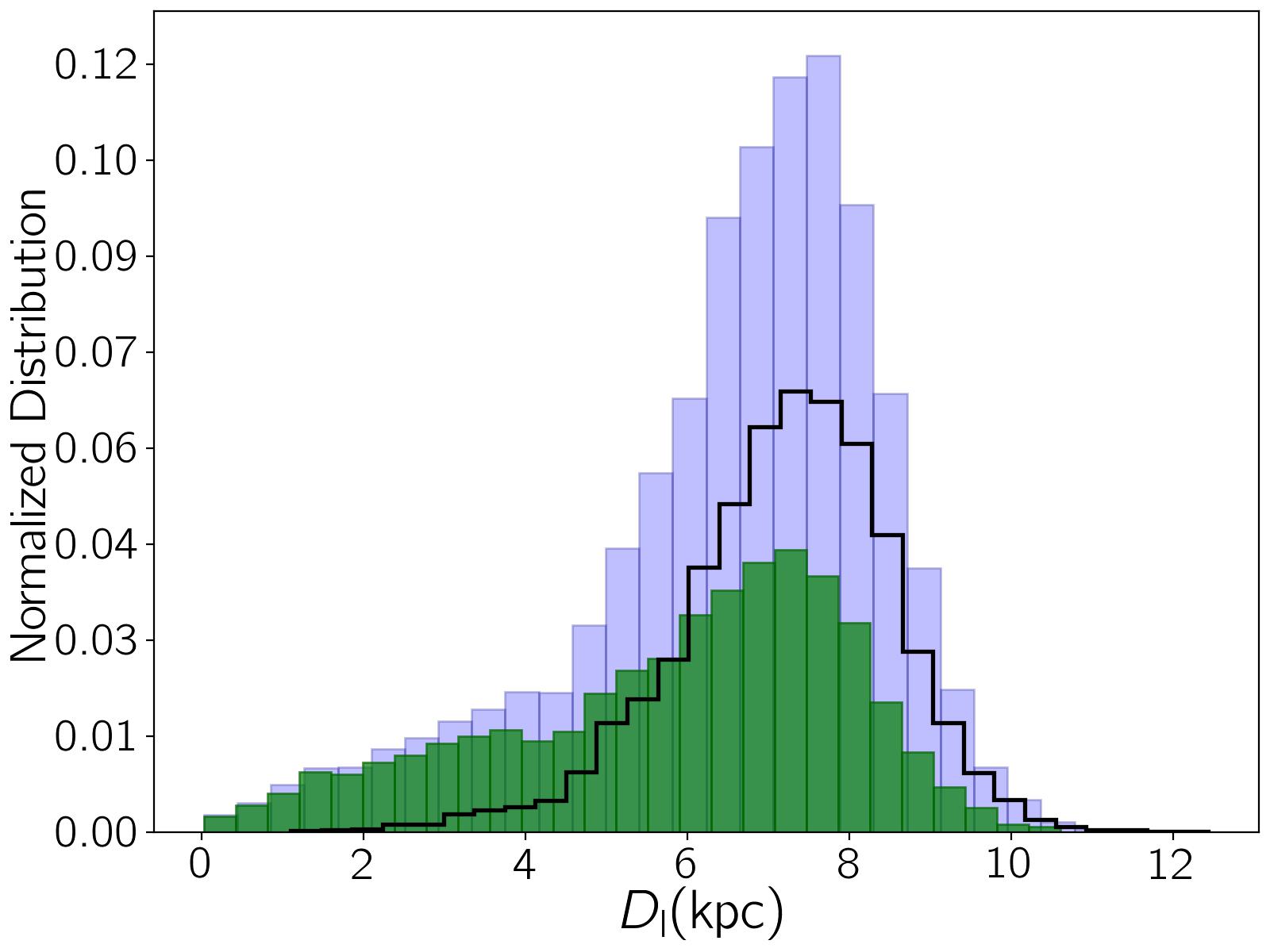}
\includegraphics[width=0.33\textwidth]{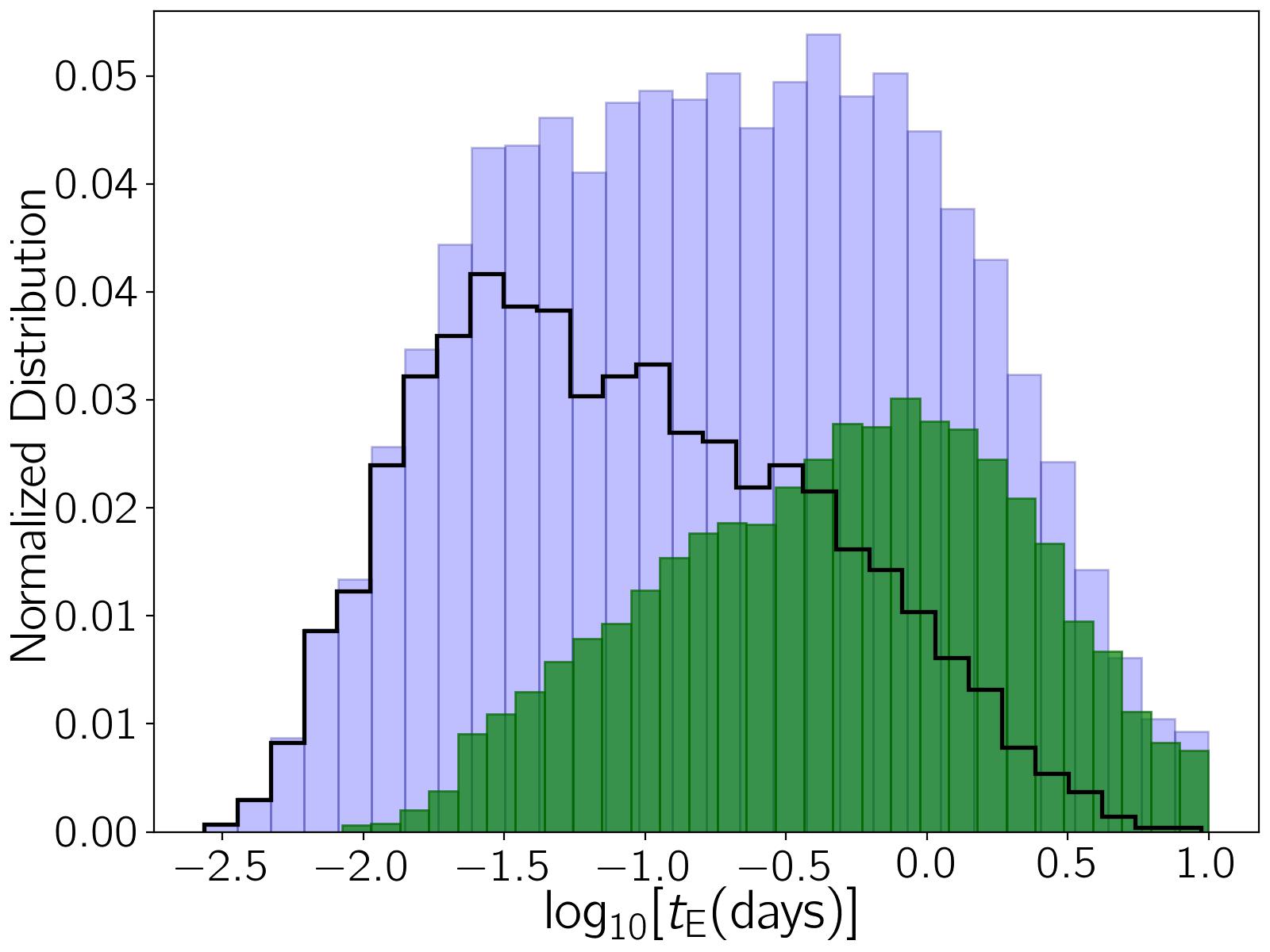}
\includegraphics[width=0.33\textwidth]{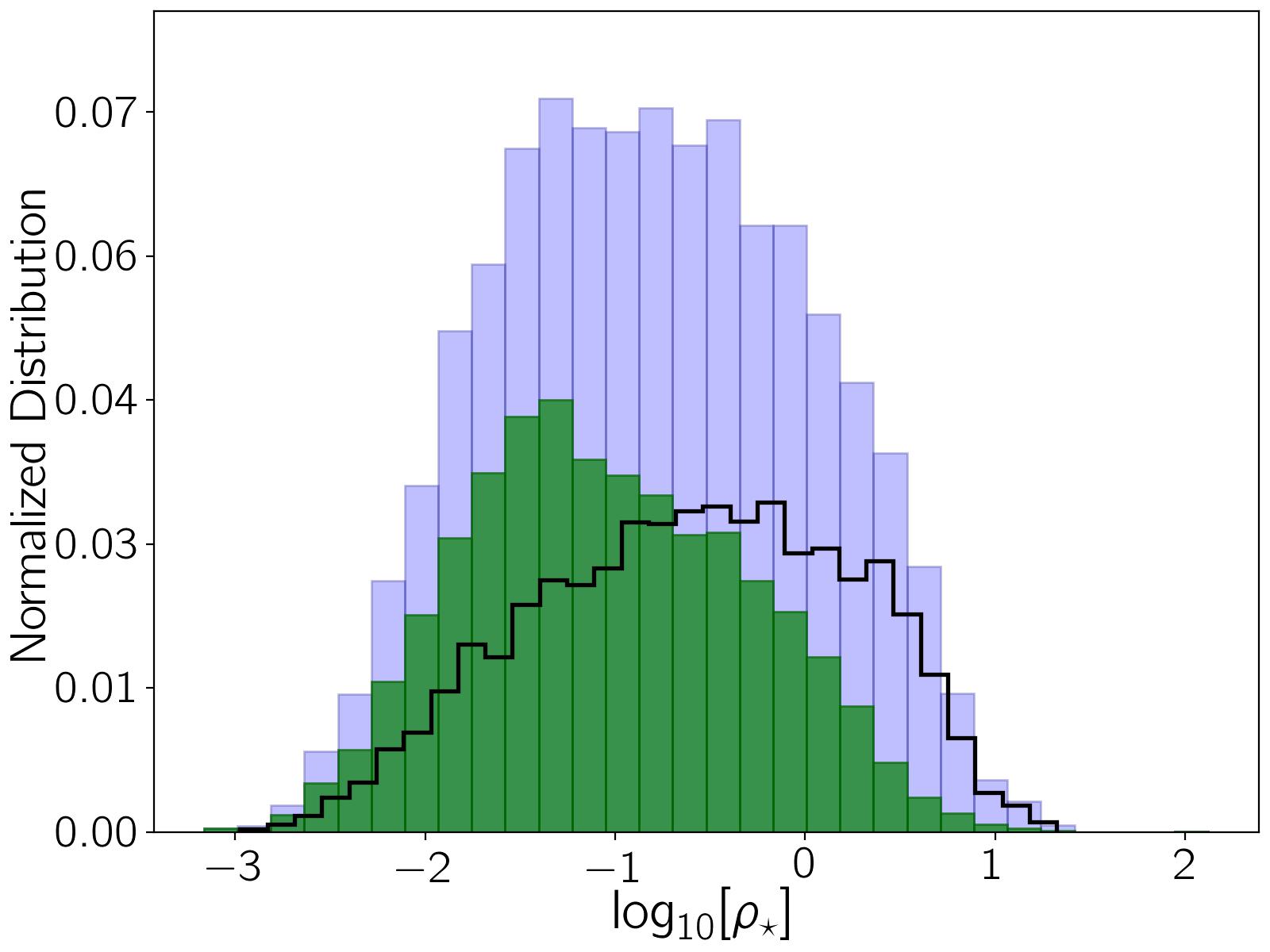}
\includegraphics[width=0.33\textwidth]{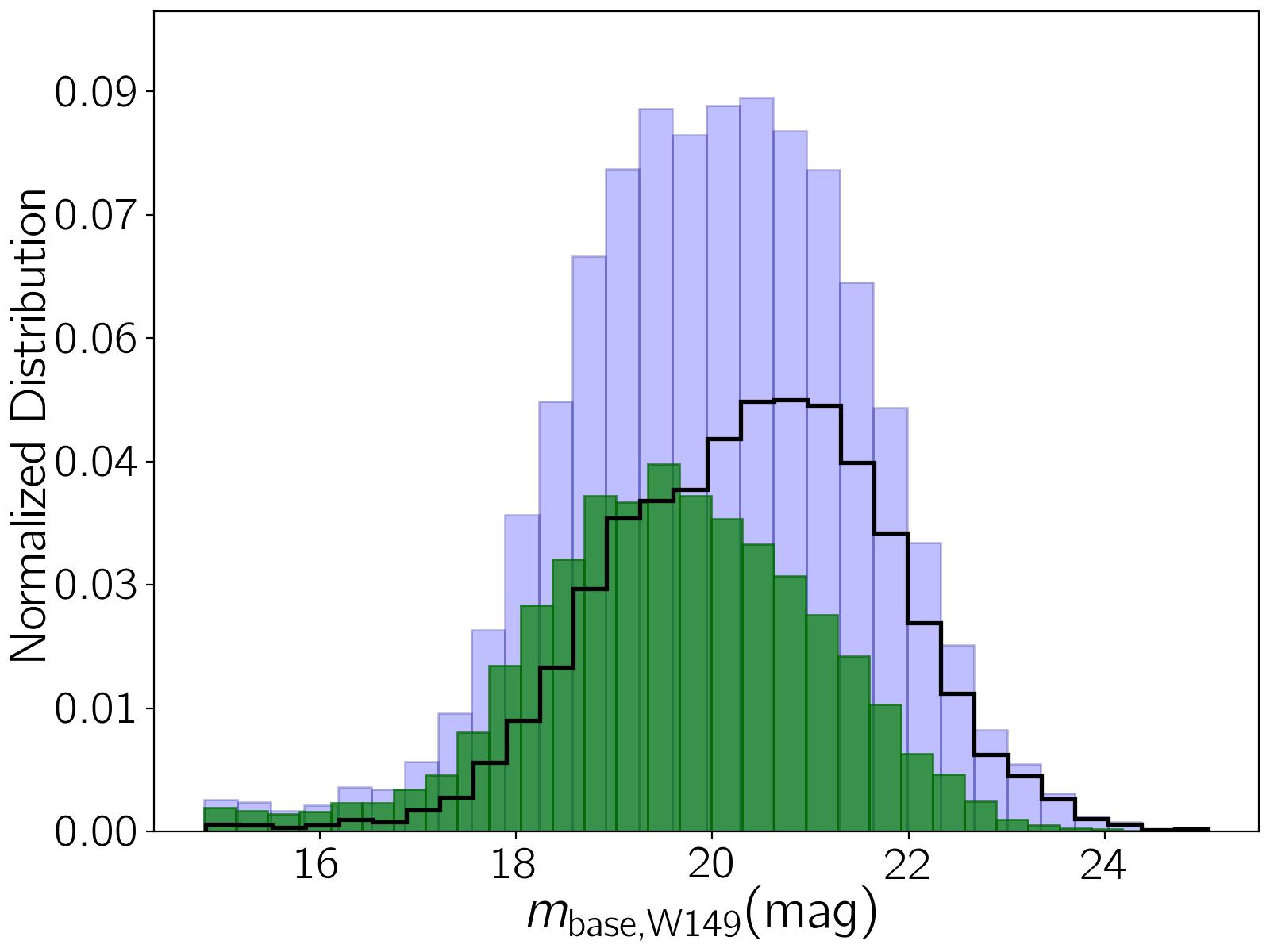}
\includegraphics[width=0.33\textwidth]{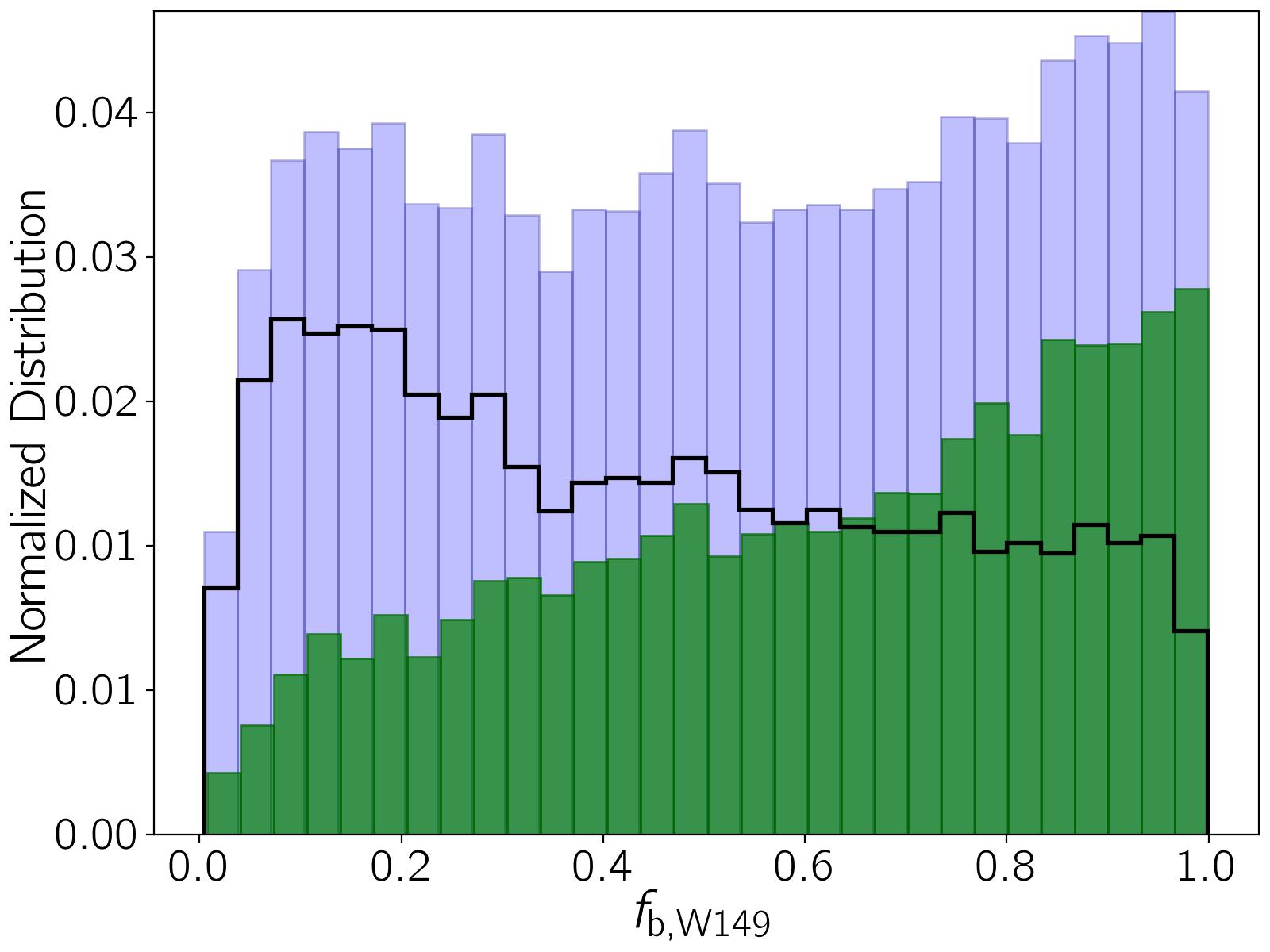}
\includegraphics[width=0.335\textwidth]{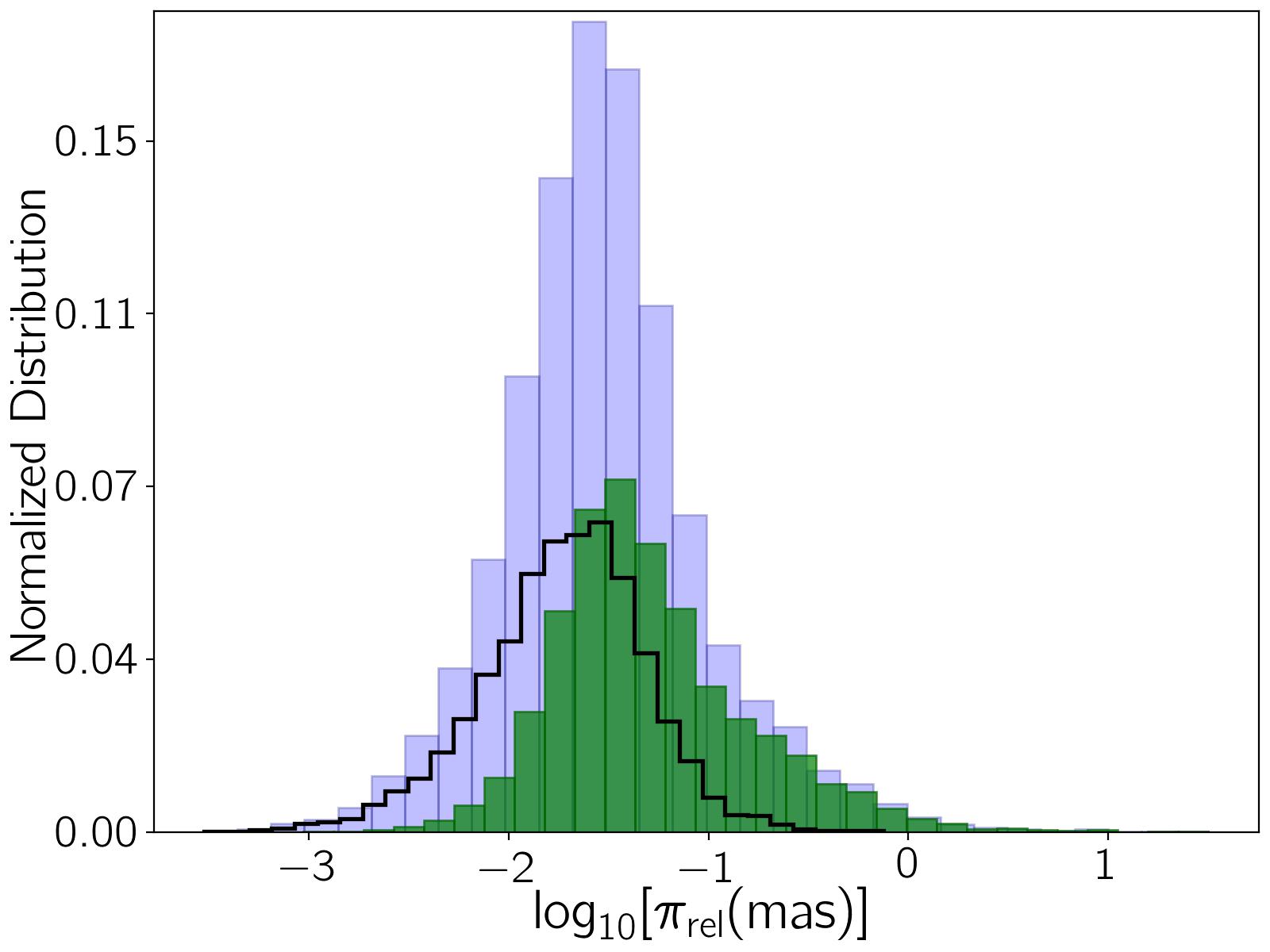}
\includegraphics[width=0.33\textwidth]{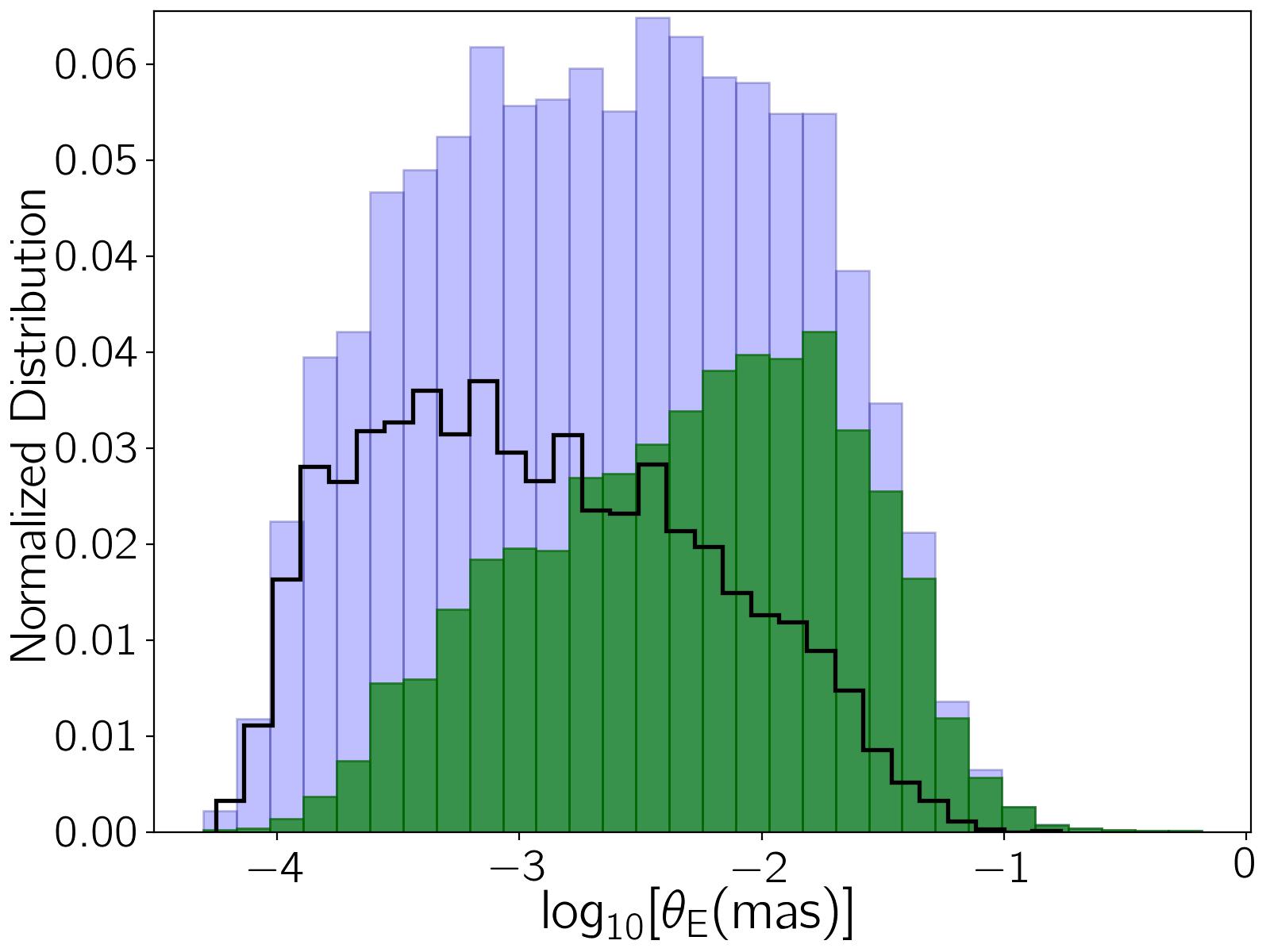}
\includegraphics[width=0.335\textwidth]{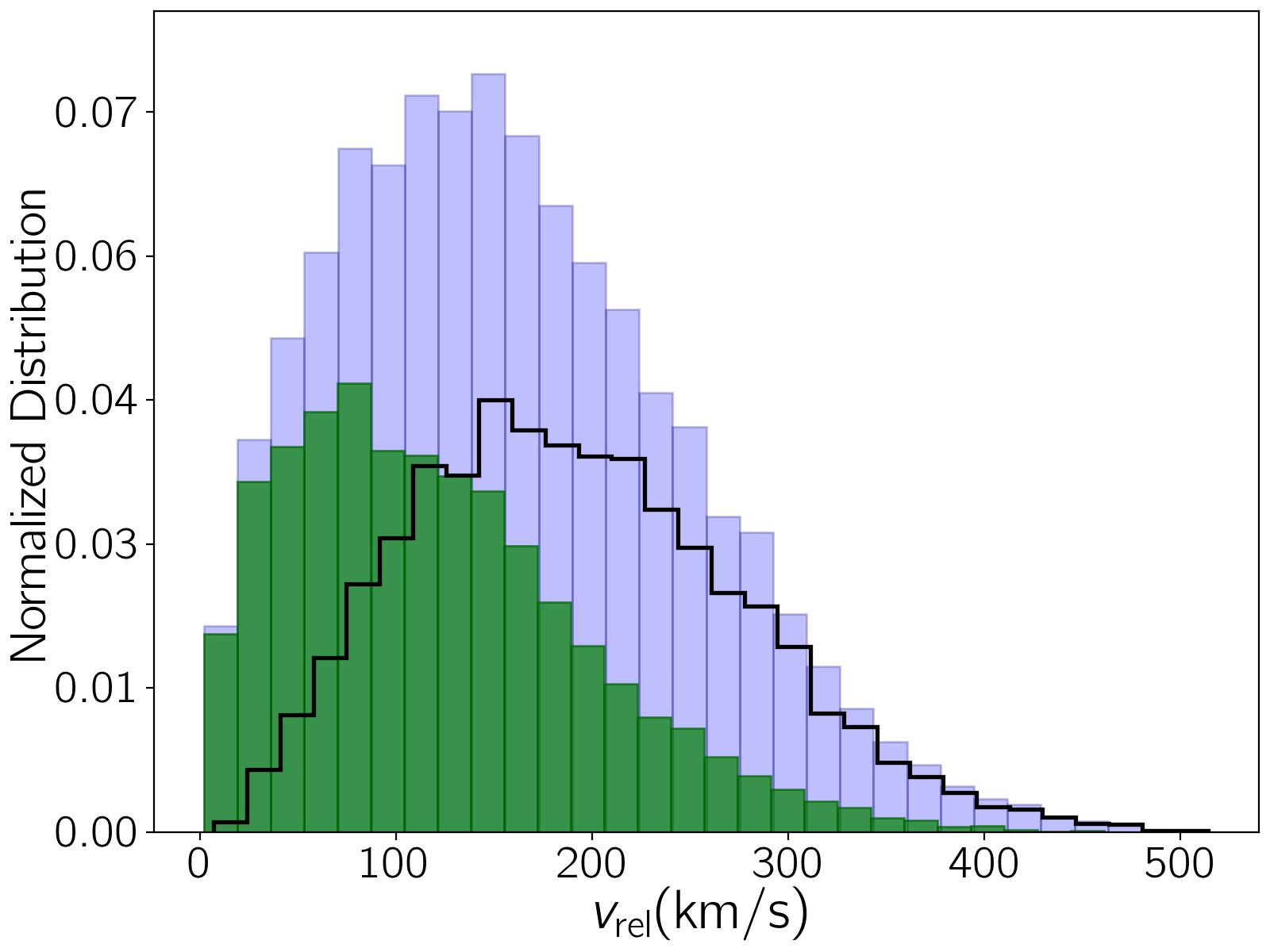}
\caption{The light purple histograms are the normalized distributions of some parameters due to all simulated microlensing events which are described in Section \ref{sec2}. The normalized distributions of the events whose light curves are (and are not) affected with parallax effects are shown with green (and black step) histograms.}\label{fig2}
\end{figure*}

\begin{deluxetable*}{c c c c c c c c c c}
\tablecolumns{10}
\centering
\tablewidth{0.99\textwidth}\tabletypesize\footnotesize
\tablecaption{Tha average values from the parameters whose distributions are shown in Figure \ref{fig2}. \label{tab1}}
\tablehead{\colhead{$\rm{Parameters}\big/\rm{from}~\rm{events}$} & \colhead{$\log_{10}[\overline{M_{\rm l}}]$}&\colhead{$\overline{D_{\rm l}}$}&\colhead{$\log_{10}[\overline{t_{\rm E}}]$}&\colhead{$\log_{10}[\overline{\rho_{\star}}]$}& \colhead{$\overline{m_{\rm{base}}}$} &\colhead{$\overline{f_{\rm{bl}}}$} & \colhead{$\log_{10}[\overline{\pi_{\rm{rel}}}]$} & \colhead{$\log_{10}[\overline{\theta_{\rm E}}]$}&\colhead{$\overline{v_{\rm{rel}}}$}  \\
	&$M_{\oplus}$&$\rm{kpc}$&$\rm{days}$ & &$\rm{mag}$ &  & $\rm{mas}$& $\rm{mas}$ & $\rm{km}/s$ }
\startdata
$\rm{All}~\rm{events}$ &$2.33$ & $6.58$ &$-0.16$&$-0.14$&$20.02$&$0.53$&$-1.12$&$-2.09$&$159.10$\\
$\rm{Parallax}-\rm{affected}~\rm{events}$&$2.46$ & $5.86$ &$0.07$&$-0.40$&$19.60$&$0.61$&$-0.88$&$-1.90$&$120.91$\\
$\rm{Not}~\rm{affected}~\rm{events}$&$2.16$ & $7.19$ &$-0.55$&$0.01$&$20.38$&$0.45$&$-1.55$&$-2.39$&$192.05$\\
\enddata
\end{deluxetable*}

\section{Monte Carlo simulation: Microlensing due to FFPs by \wfirst}\label{sec2}
In this section we do a Monte Carlo simulation from microlensing events due to FFPs that can be detected during the \wfirst\ microlensing Survey. We assume that the \wfirst\ telescope is taking data from these events during one $62$-day observing season in $7$ lines of sight toward the Galactic bugle and there are no follow-up observations for these events.  

The details of these simulations can be found in the previous papers \citep[see, e. g., ][]{2021MNRASsajadianhZP, 2023AJsajadianast, 2023MNRASsajadiansang}, so we briefly explain them here. To generate a microlensing event, we should choose (i) a source star in a given line of sight, and (ii) a lens object, and then (iii) determine the lensing parameters. We specify the distances of source stars ($D_{\rm s}$) based on the overall mass density versus the distance from the observer in a given line of sight. Then, we determine the photometry properties of the source stars using the Besan\c{c}on model, and according to the Galactic structure to which the source star belong \citep{Robin2003,Robin2012}. In the next step, we indicate the lens distance from the observer ($D_{\rm l}$) using the microlensing event rate function. We limit the masses of lenses to the mass range $M_{\rm l} \in [0.01 M_{\oplus}, 15 M_{\rm J}]$ and select them using a log-uniform mass function $dN/dM_{\rm l} \propto M_{\rm l}^{-1}$, in the same way with \citet{2020AJjohnson,2021sajadian}. We determine the lens and source dispersion velocity components using the normal distributions whose widths given by \citet{Robin2003}. In the short-duration microlensing events due to FFPs, the finite-source effect \citep{1994wittmoa} is considerable, since the angular source radius $\theta_{\star}$ normalized to the angular Einstein radius is $\rho_{\star}=\theta_{\star} \big/\theta_{\rm E} \propto M_{\rm l}^{-1/2}$. Therefore, we determine the microlensing magnification factor by considering finite-source size and using the RT-model \citep{2010MNRASBozza,2018MNRASBozza}. For these events we consider the parallax effect based on the formalism explained in the previous section. Since we aim to focus on the events affected by the parallax whereas this effect is unrecognizable while modeling (or when this effect should be ignored while inferring the lensing parameters), we exclude the events have $t_{\rm E}>10$ days. However, in the simulation $99\%$ of simulated events due to FFPs with a log-uniform mass function had $t_{\rm E}$ less than $10$ days.  

\begin{figure*}
\centering
\includegraphics[width=0.49\textwidth]{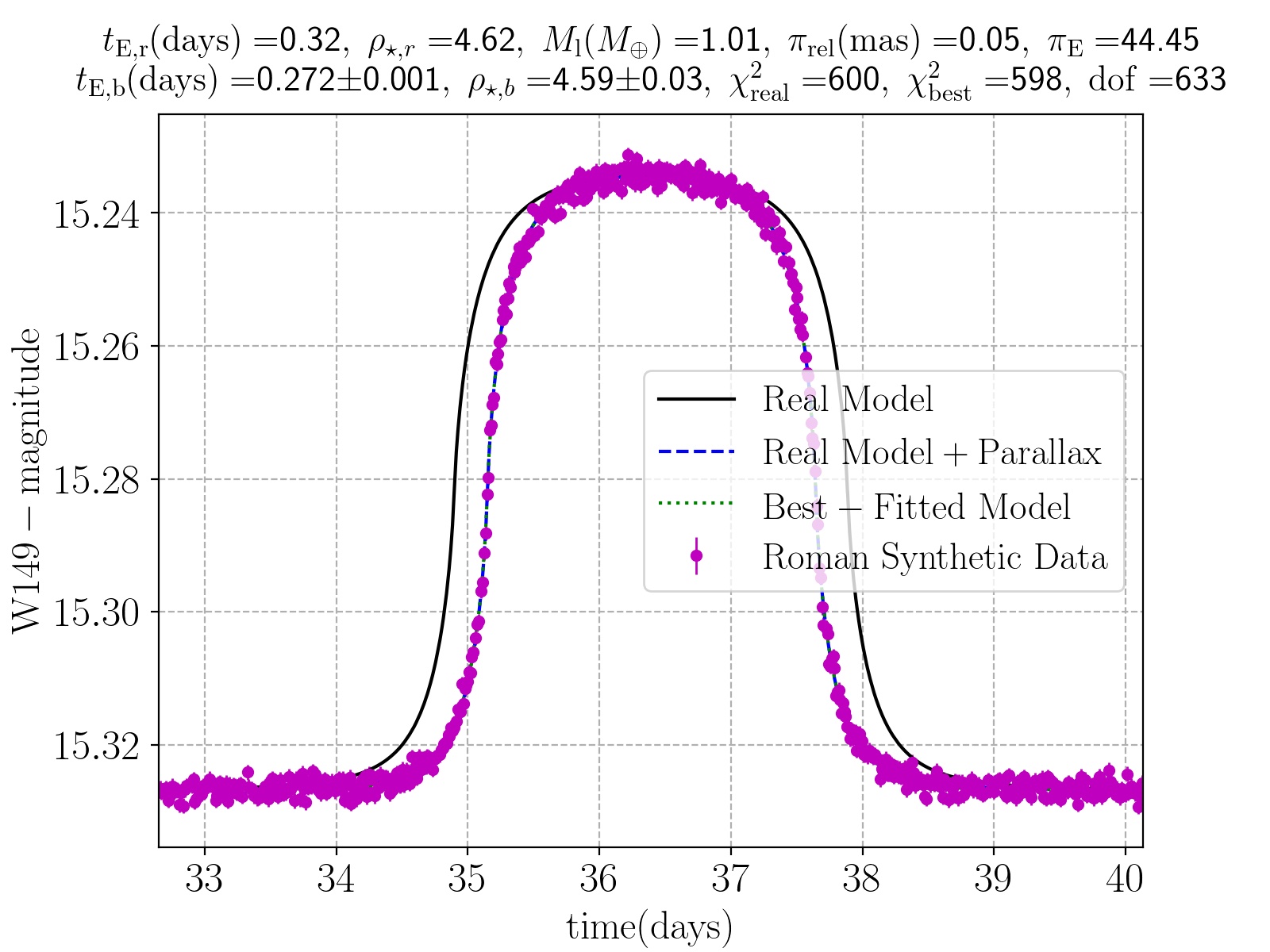}
\includegraphics[width=0.49\textwidth]{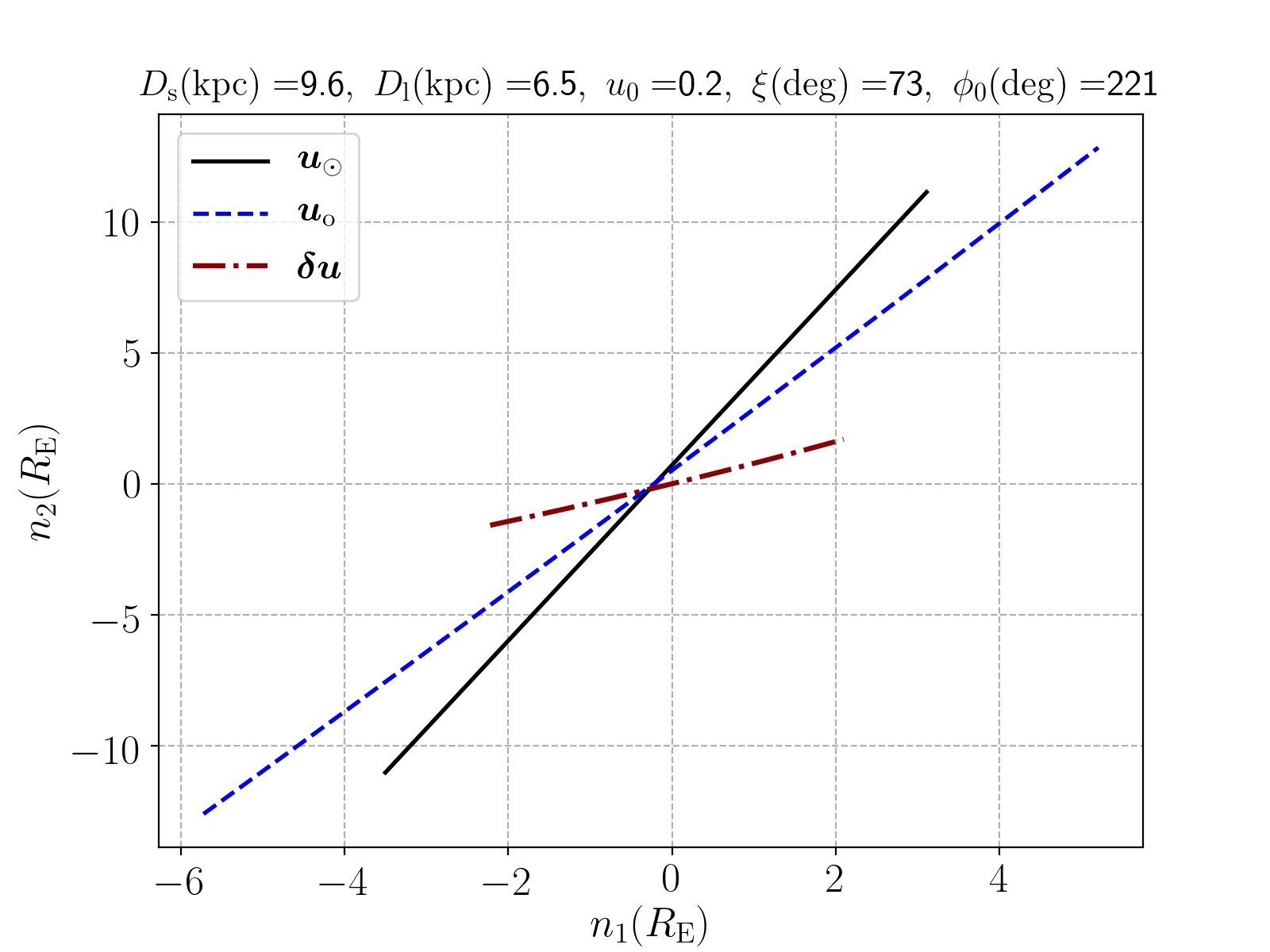}
\includegraphics[width=0.49\textwidth]{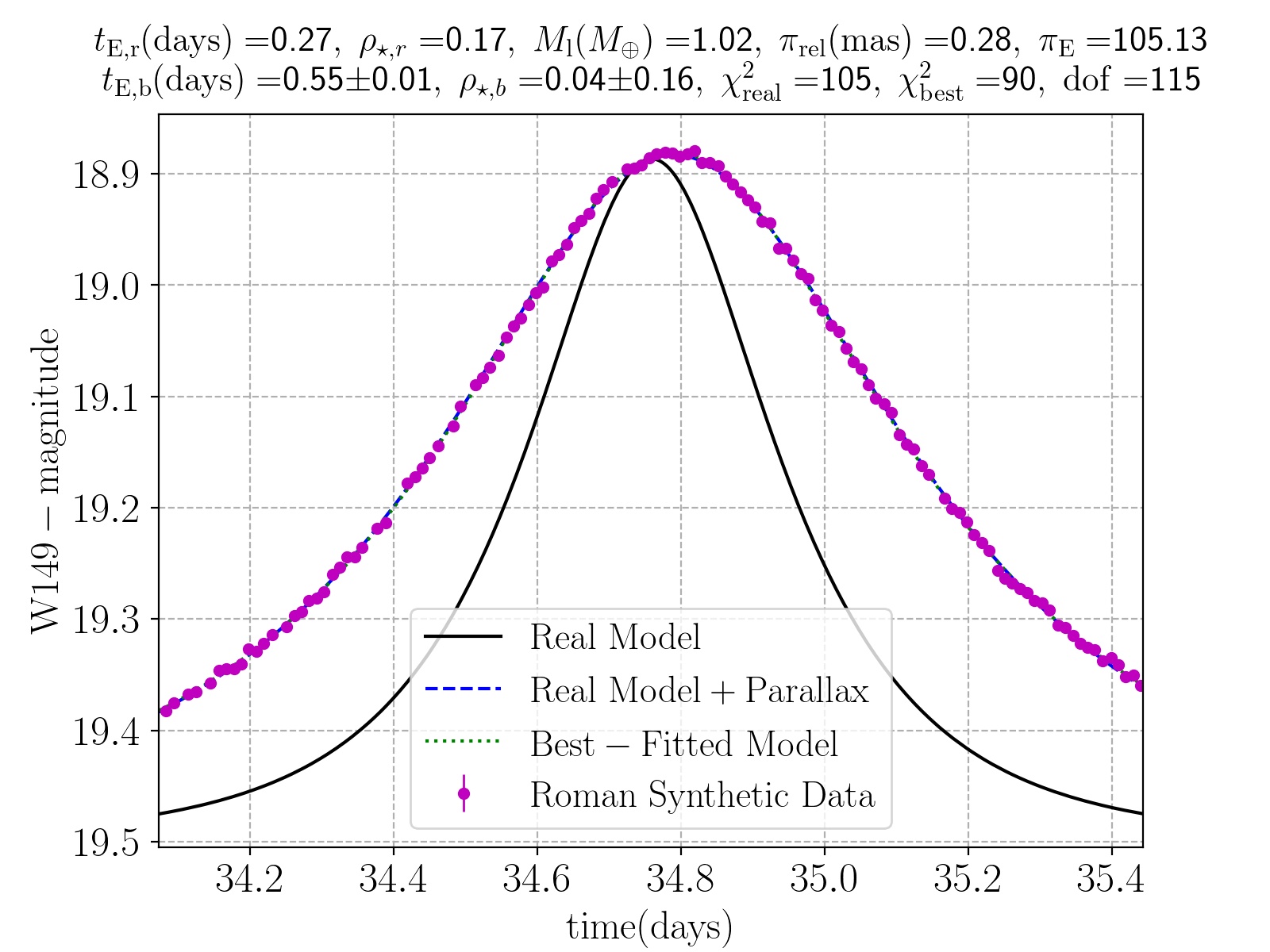}
\includegraphics[width=0.49\textwidth]{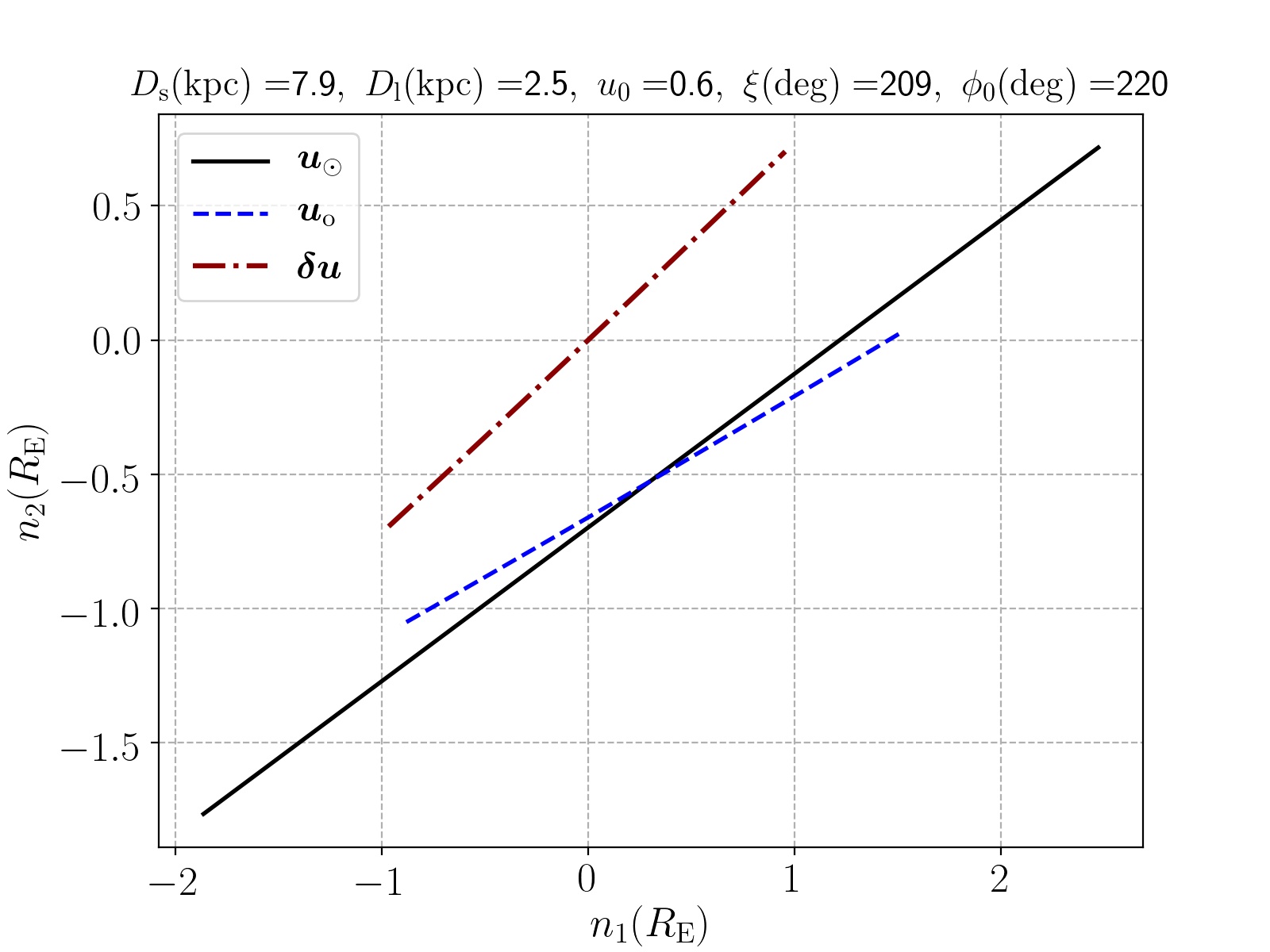}
\caption{Similar to Figure \ref{fig1}. In these two short-duration microlensing light curves the parallax changes their widths significantly. The index $b$ for $t_{\rm E}$ and $\rho_{\star}$ refers to their values in the best-fitted simple microlensing models. The best-fitted light curves are depicted by green dotted curves. In the titles 'dof' refers to the degree of freedom.}\label{fig3}
\end{figure*}

\begin{figure*}
\centering
\includegraphics[width=0.49\textwidth]{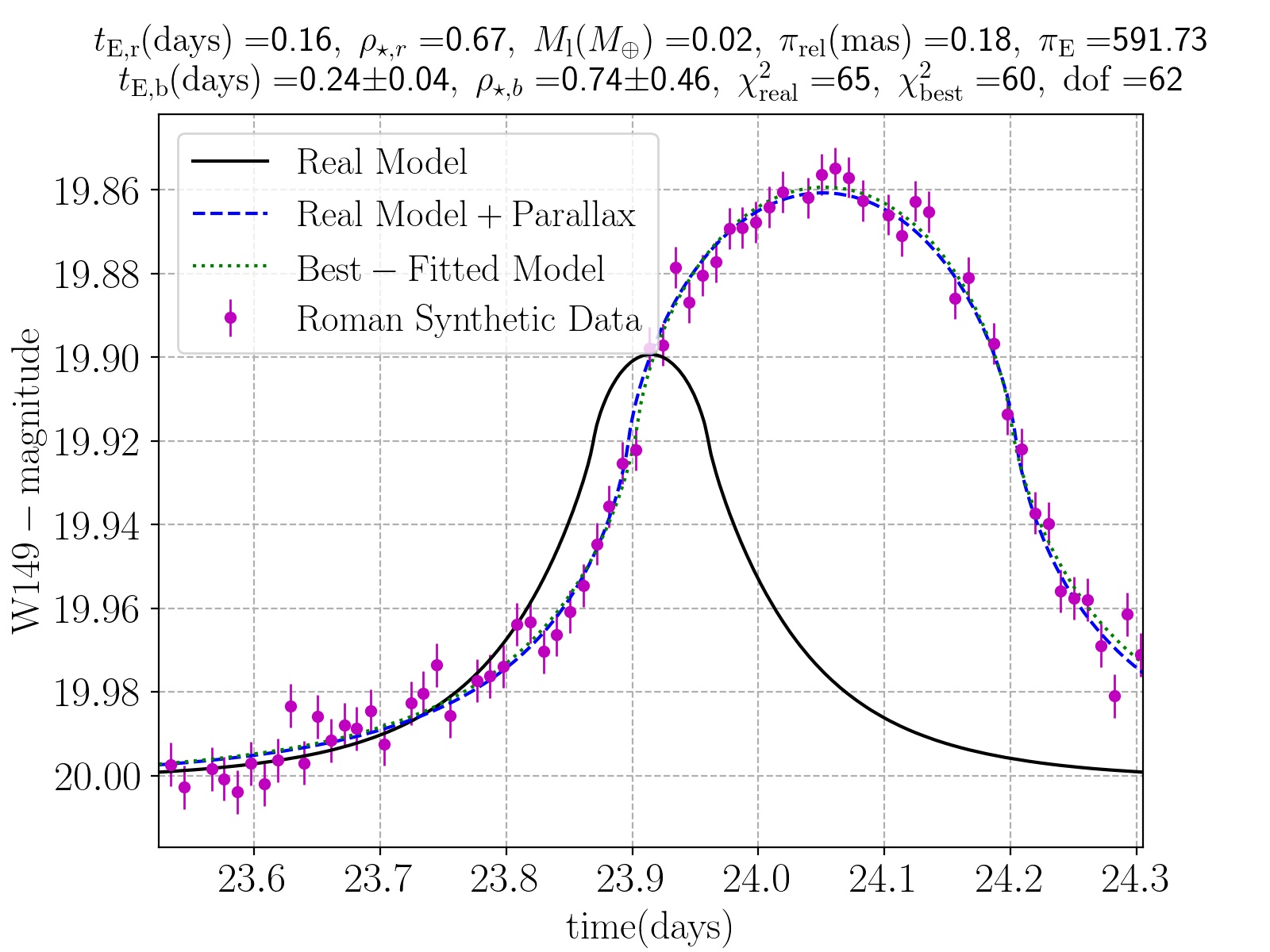}
\includegraphics[width=0.49\textwidth]{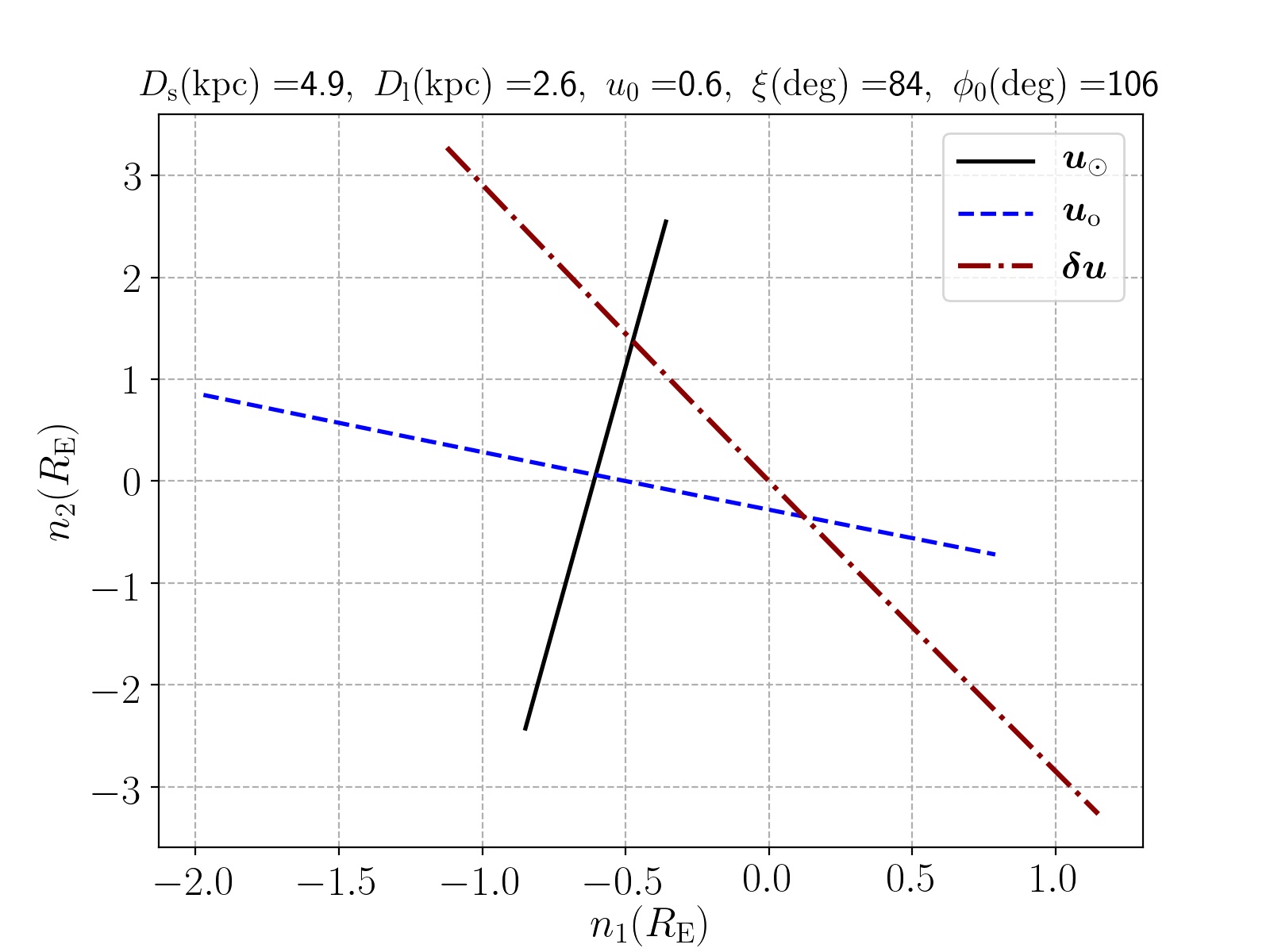}
\includegraphics[width=0.49\textwidth]{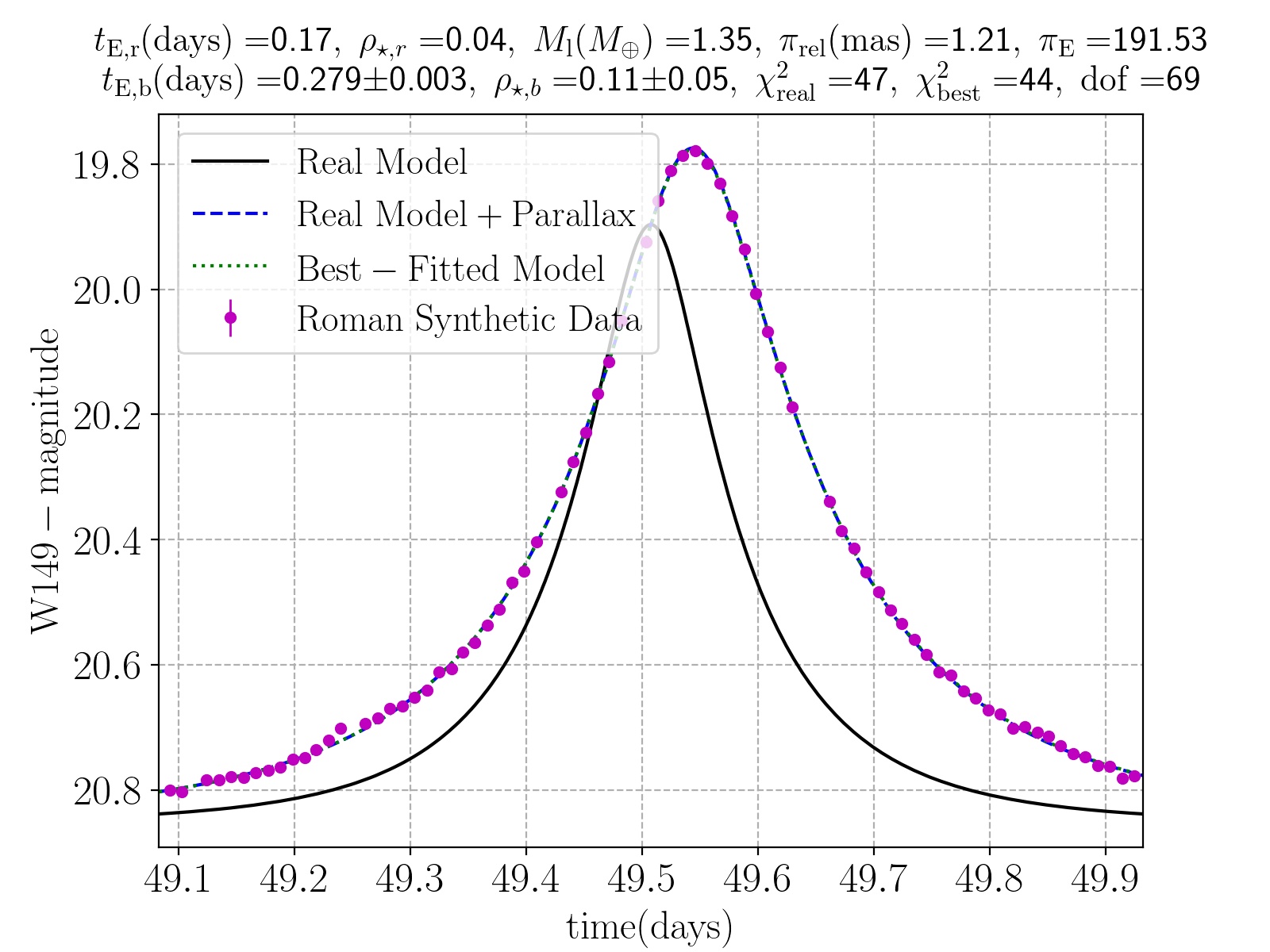}
\includegraphics[width=0.49\textwidth]{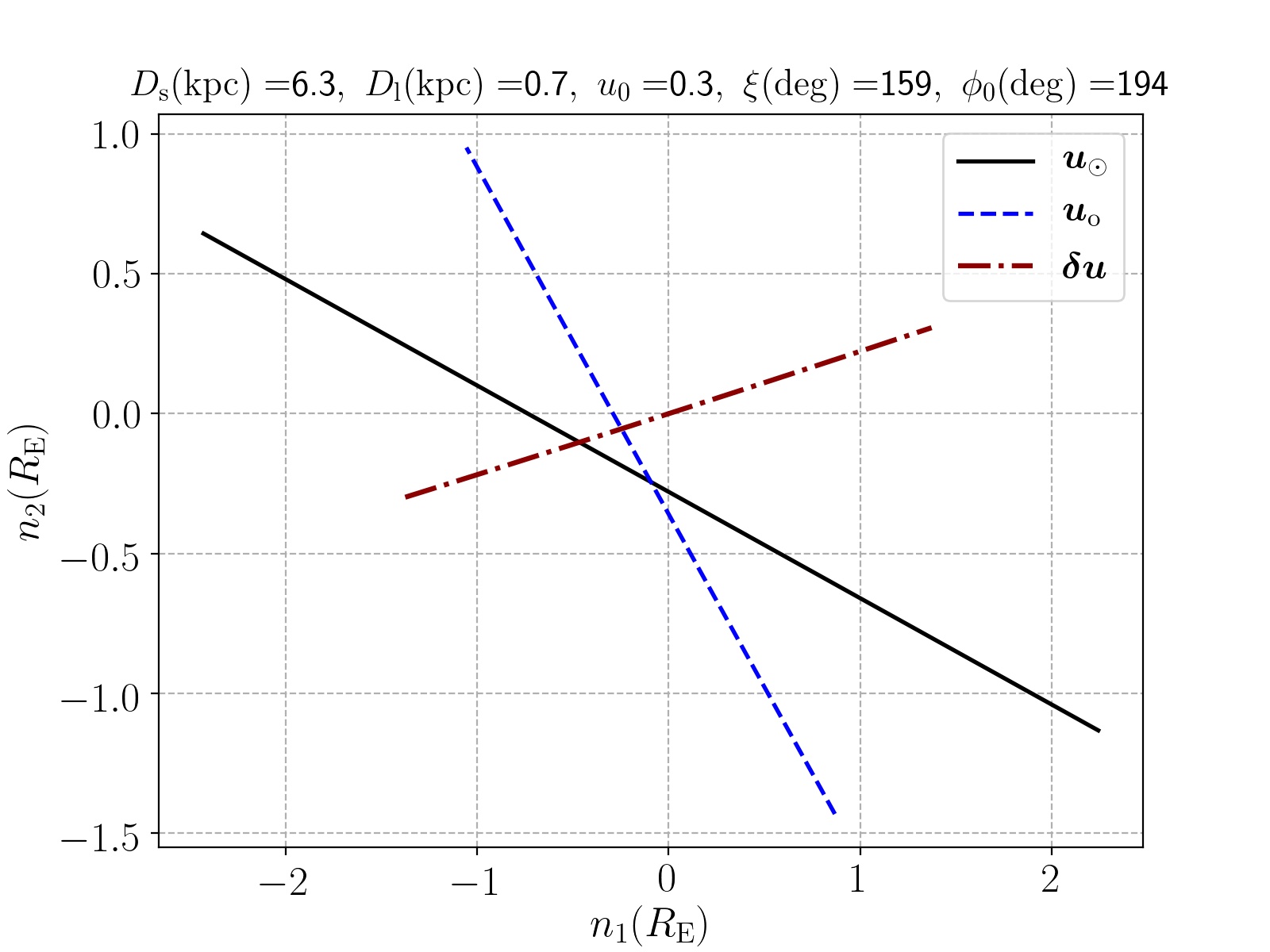}
\caption{Similar to Figure \ref{fig3}. In these two microlensing events the parallax-induced deviations are significant, because in these events the lens object is close to the observer which results large $\pi_{\rm{rel}}$ values. }\label{fig4}
\end{figure*}
In the next step, for each light curve we produce the synthetic data points taken by the \wfirst\ telescope. We consider an observing season with the duration $T_{\rm{obs}}=62$ day, and uniformly choose $t_{0}$s from the range $[0,~T_{\rm{obs}}]$. For each data point we determine the error bar in the magnification factor using $\sigma_{\rm{A}}=A \big|1- 10^{-0.4~\sigma_{\rm{m}}}\big|$, where $A$ is the magnification factor. $\sigma_{\rm m}$ is the \wfirst\ photometric error which is a function of the apparent magnitude in the \wfirst\ filter (W149) \citep[Fig. 4 in][]{2019ApJPenny}. The cadence between data is fixed to $15.16$ min. 

In the simulation we determine the number of blending stars by integrating over the Galactic number density in a given line of sight to determine the projected surface one  \citep[Eq. 2, 3 in ][]{2019ApJsajadian}. This surface density indicates the number of blending stars whose lights enter the source PSF (Point Spread Function). The blending in the lensing formalism is usually evaluated by the factor $f_{\rm bl}$ which is the ratio of the source flux (itself) to total baseline flux enters the source PSF. The higher $f_{\rm bl}$ value, the less number of blending stars. We perform this Monte Carlo simulation, and generate 9646 microlensing events due to FFPs with $t_{\rm E}<10$ days which potentially are discernible by the \wfirst\ telescope. The detectability criteria were explained in \citet{2021sajadian}. 

For some of these events, the parallax effect does not change the light curve shape and so ignoring the parallax effect for these events does not alter the inferred lensing parameters significantly. To extract these events we consider a criterion, i.e., the difference between $\chi^{2}_{\rm{real}}$ and $\chi^{2}_{\rm{without}}$ values from fitting the real model with parallax and the real model without parallax (respectively) be less than 100, i.e., $\Delta \chi^{2}=\big| \chi^{2}_{\rm{real}}-\chi^{2}_{\rm{without}}\big|<100$. Most of these events have either very short time scales (with sparse data points), or large finite-source sizes or faint source stars with high photometric error bars which all result small $\Delta \chi^{2}$ values. Additionally, when $D_{\rm l}\simeq D_{\rm s}$ which results very small $\pi_{\rm{rel}}$ values, the parallax amplitude decreases, and this effect will not change the microlensing light curve.  

\noindent Two examples of these events can be found in Figure \ref{fig1}. In this figure, for two light curves (shown in the left panels), their lens-source relative trajectories with (blue dashed line $\boldsymbol{u}_{\rm o}$) and without (black solid line $\boldsymbol{u}_{\odot}$) parallax effects are represented in their right panels. The parallax-induced deviations in the lens-source relative trajectories $\boldsymbol{\delta u}$ are depicted with the dark-red dot-dashed lines. The lensing parameters used to these light curves are mentioned at the top of plots. In these light curves high photometric errors and low number of data points cause the parallax-induced deviations to be undetectable.   

$46.3\%$ of all simulated events have $\Delta \chi^{2}>100$ and $t_{\rm E}<10$ days. For these events the parallax effect is invisible, but changes their observed light curves considerably. 

In Figure \ref{fig2}, we show the normalized distributions of some parameters from simulated microlensing events due to FFPs with light purple color. To study what kinds of simulated events are more affected with the parallax effect, in these plots the distributions of events with $\Delta \chi^{2}>100$ (and $\Delta \chi^{2}<100$) are plotted with green (and black step) histograms. To better compare these distributions, in Table \ref{tab1} we report the average values of the parameters whose distributions are plotted in Figure \ref{fig2}. In this table three rows (from top to bottom) contain the average values due to all simulated events (with light purple histograms), parallax-affected events (green histograms) and not affected ones (black step ones), respectively.  

Accordingly, most of events in which $D_{\rm l}\lesssim3$ kpc (or $\pi_{\rm{rel}}\gtrsim 0.2$ mas), and $t_{\rm E}\gtrsim3$ days (which have more number of data points) are affected by parallax with $\Delta\chi^{2}>100$. Also, the events due to faint or highly blended source stars, or the ones with $\rho_{\star}\gtrsim 3$ are less affected with parallax (e.g., the top event in Fig. \ref{fig1}). In fact, when the normalized source radius is large, the magnification factor is low as it is estimated by $1+2\big/\rho_{\star}^{2}$ \citep{1996Gouldfinite,2003ApJAgol}, so variations in the lens-source relative trajectory does not change the light curve much.

In the next section, we focus on the microlensing events with $\Delta \chi^{2}>100$ and study the parallax-induced deviations in their lensing parameters by finding the best-fitted simple Paczy\'nski light curves for these events.

\subsection{Fitting simple Paczy\'nski microlensing models}
For each of the simulated microlensing events due to FFPs with $t_{\rm E}<10$ days and $\Delta \chi^{2}>100$, we find the best-fitted simple Paczy\'nski microlensing models using the python-based package \texttt{emcee}\footnote{\url{https://emcee.readthedocs.io/en/stable/}}\citep{2013PASPForeman}. For fitting a simple microlensing model there are $5$ parameters which are $t_{\rm E}$, $\rho_{\star}$, $t_{0}$, $u_{0}$, and $f_{\rm b}$. We exclude the apparent baseline magnitude due to all blending stars in the W149 filter, $m_{\rm{base}}$, because the parallax does not change this parameter. For finding the best-fitted microlensing models with 5 free parameters, 40 chains (i.e., Nwalkers$=40$ in the \texttt{emcee} code) and each chain with $10000$ steps were sufficient. We ignore the limb-darkening effect for source stars, and simulate synthetic data points for light curves with $\rho_{\star}<1$ in the time interval $[-2.5~t_{\rm E}+ t_{0},~2.5~t_{\rm E}+ t_{0}]$, and the others in the time interval $[-2.5~t_{\star}+ t_{0},~2.5~t_{\star}+ t_{0}]$, where $t_{\star}=t_{\rm E}\times \rho_{\star}$. 

\begin{deluxetable}{c c c c c c c}
\tablecolumns{7}
\centering
\tablewidth{0.9\textwidth}\tabletypesize\footnotesize  
\tablecaption{The statistical results from the Monte Carlo simulation. \label{tab2}}
\tablehead{\colhead{$ $}& \colhead{$\delta t_{\rm E}$}& \colhead{$\delta \rho_{\star}$} & \colhead{$\delta u_{0}$}& \colhead{$\delta t_{0}$} &\colhead{$\delta f_{\rm bl}$}  &  \colhead{$\delta t_{\rm E}$~and~$\delta \rho_{\star}$}} 
\startdata  
$> 0.02$&$73.7$&  $83.6$  &  $63.2$  &  $3.6$  &  $52.1$&  $62.6$\\
$> 0.05$&$46.7$&  $75.6$  &  $43.3$  &  $1.5$  &  $31.8$&  $36.8$\\
$> 0.1$&$26.8$&  $68.8$  &  $29.4$  &  $0.7$  &  $18.6$&  $19.8$\\
$> 0.3$&$8.0$  &  $56.9$  &  $14.7$  &  $0.2$  &  $4.8$  &  $4.9$\\
$> 0.5$&$3.8$  &  $48.7$  &  $10.1$  &  $0.1$  &  $1.9$  &  $1.9$\\
$> 0.7$&$1.9$  &  $41.3$  &  $7.5$  &  $0.1$  &  $0.8$  &  $0.8$\\
\enddata
\tablecomments{Each entrance is the fraction of simulated events with $\Delta \chi^{2}>100$ in which the given relative deviations (its column) are larger than the given threshold (its row).}
\end{deluxetable}

\begin{figure*}
\centering
\includegraphics[width=0.49\textwidth]{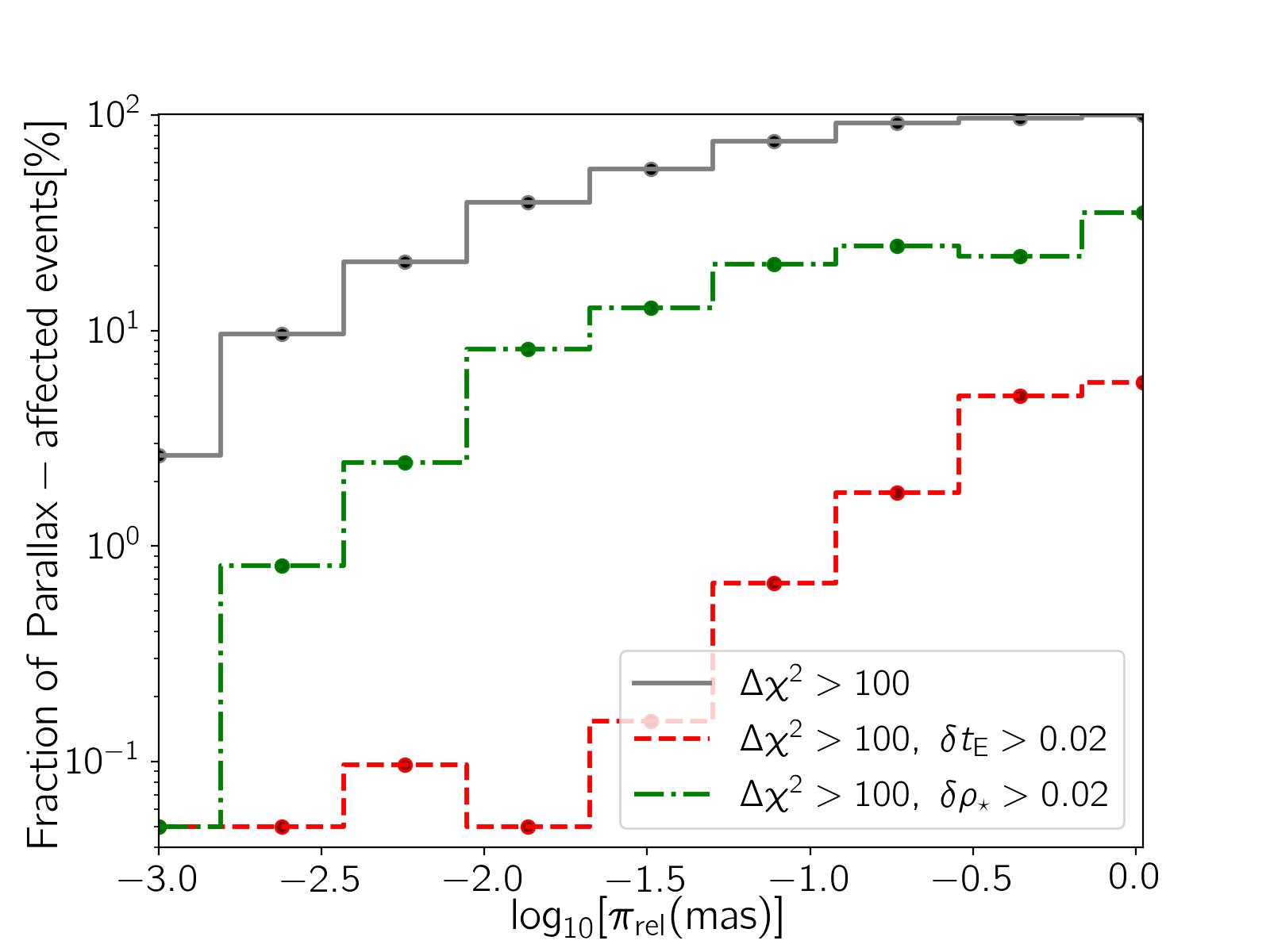}
\includegraphics[width=0.49\textwidth]{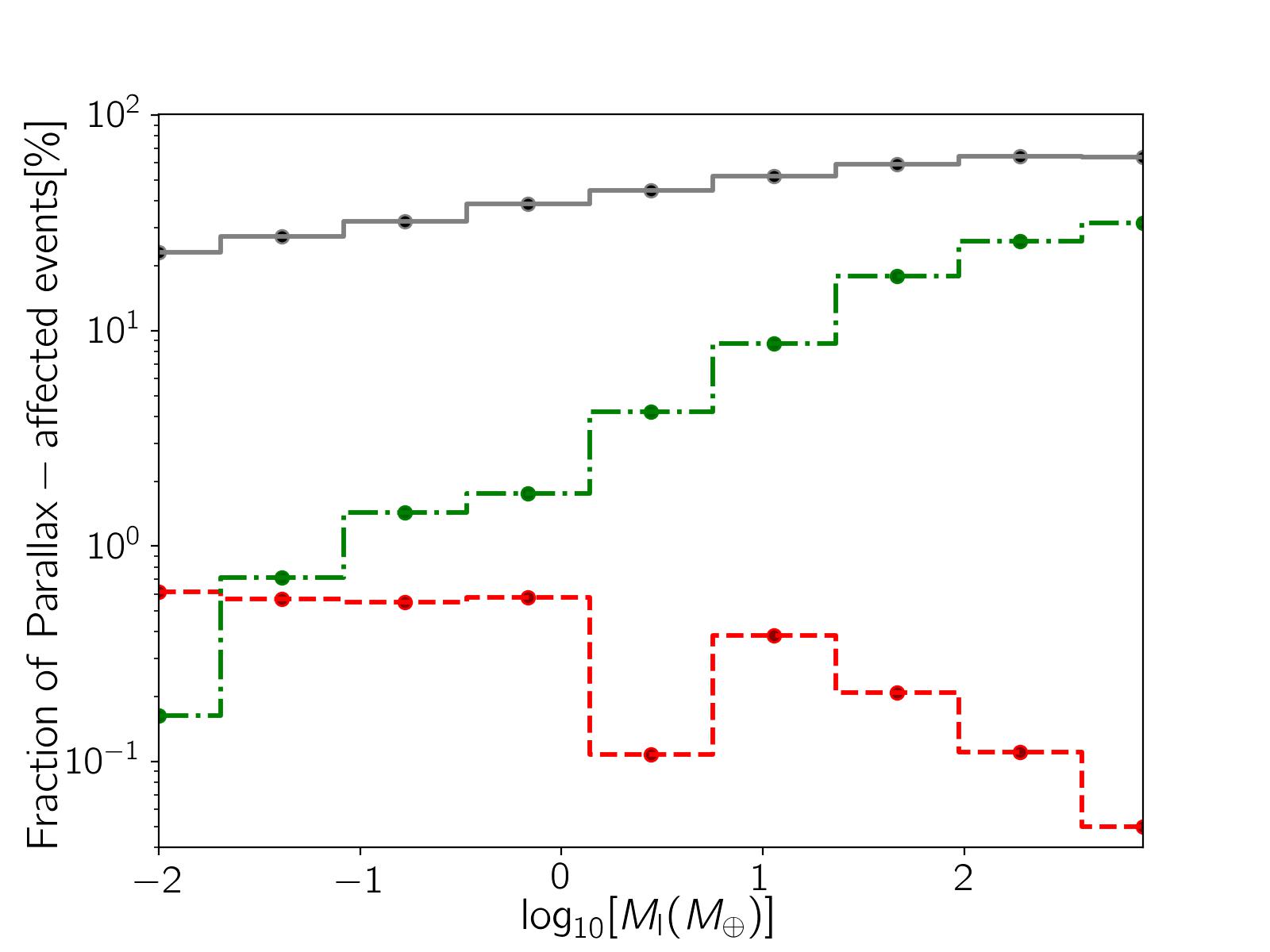}
\includegraphics[width=0.49\textwidth]{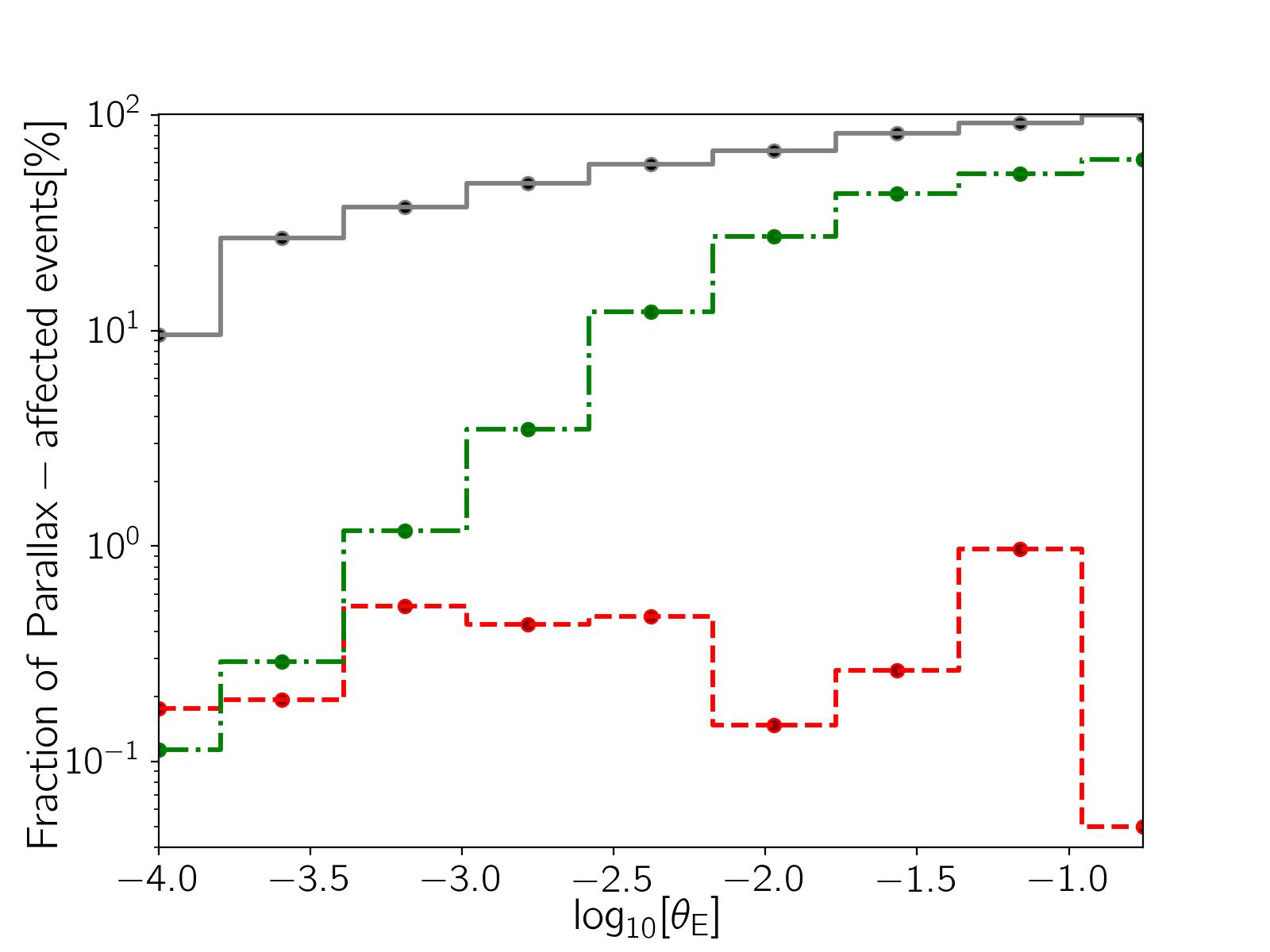}
\includegraphics[width=0.49\textwidth]{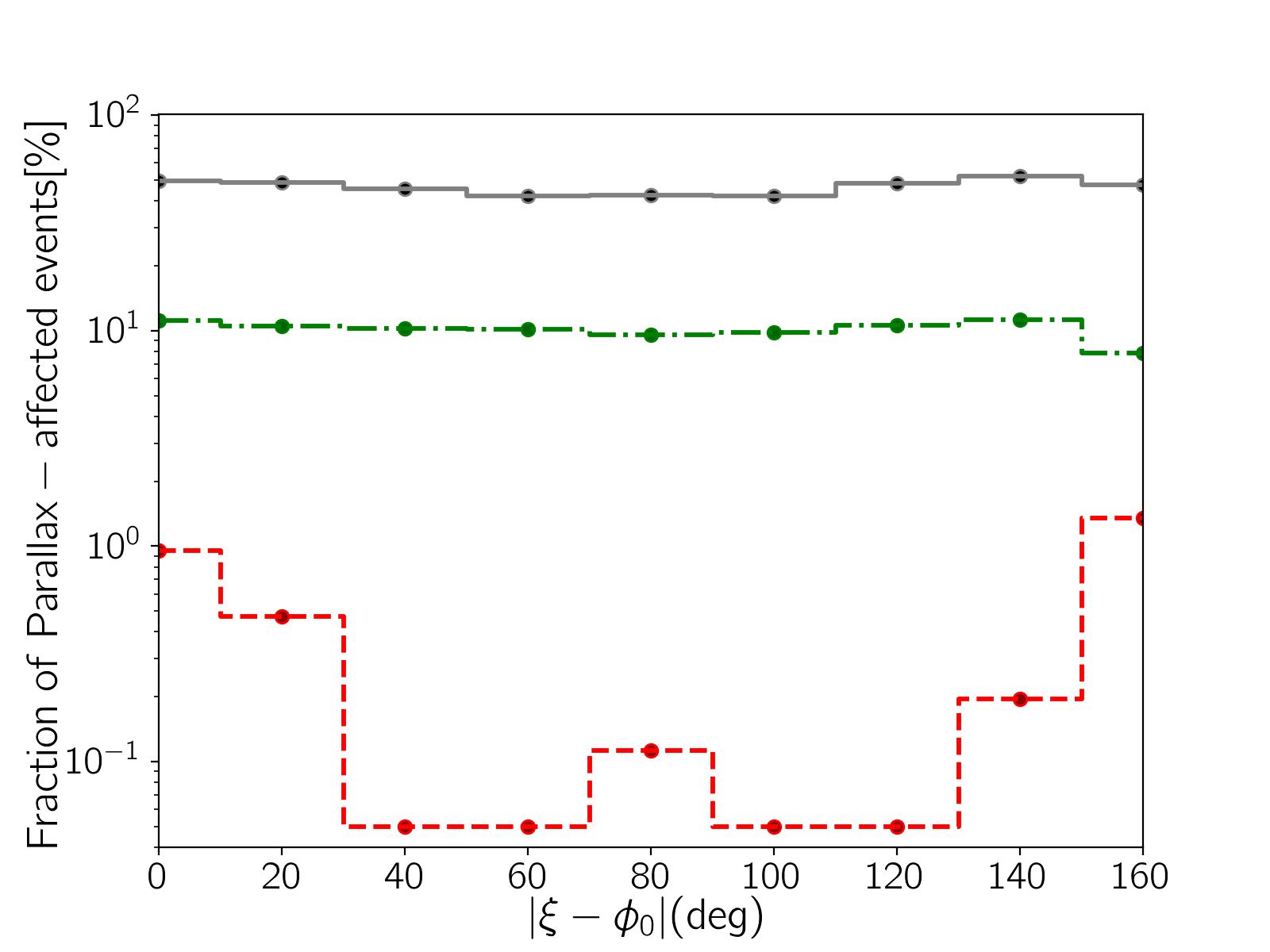}
\caption{The fraction of the simulated events due to FFPs which are affected by parallax in percent. Here, we have three definitions to extract the parallax-affected events: (i) $\Delta \chi^{2}>100$ (gray solid lines), (ii) $\Delta \chi^{2}>100$ and $\delta t_{\rm E}>0.02$ (red dashed lines), and (iii) $\Delta \chi^{2}>100$ and $\delta \rho_{\star}>0.02$ (green dot-dashed lines).}\label{fig5}
\end{figure*}

Two parameters $t_{\rm E}$ and $\rho_{\star}$ are functions of the lens mass and its distance. Hence, the parallax-induced deviations from these two parameters directly lead to misinterpretation of the lens object. From simulations, we found that in events that $\boldsymbol{u}_{\odot}$ and $\boldsymbol{\delta u}$ are parallel, i.e.,  $\big| \xi-\phi_{0}\big|\simeq 0,~180^{\circ}$ the widths of observed light curves change significantly and lead to the highest deviations in either $t_{\rm E}$ or $\rho_{\star}$. Two examples of such events are represented in Figure \ref{fig3}. In these light curves the best-fitted light curves are shown with green dotted curves. $t_{\rm E,~b}$ and $\rho_{\star,~b}$  are the Einstein crossing time and the normalized source radius due to the best-fitted models which are mentioned in the second lines of the left panels' titles. To evaluate the goodness of the fitted models, we report the $\chi^{2}_{\rm{real}}$, and $\chi^{2}_{\rm{best}}$ values due to the real, and best-fitted models (blue dashed, and green dotted curves respectively), and the number of degrees of freedom (dof) in the titles. 

As mentioned in the previous section, when $\pi_{\rm{rel}}$ is large (the lens object is close to the observer) the parallax-induced deviations in the microlensing light curves are relatively large so that most of lensing parameters due to the best-fitted models are different from real ones. Two examples of such events are represented in Figure \ref{fig4}. The statistical results from this Monte Carlo simulation are reported in the next section.   
\begin{figure}
\centering
\includegraphics[width=0.4\textwidth]{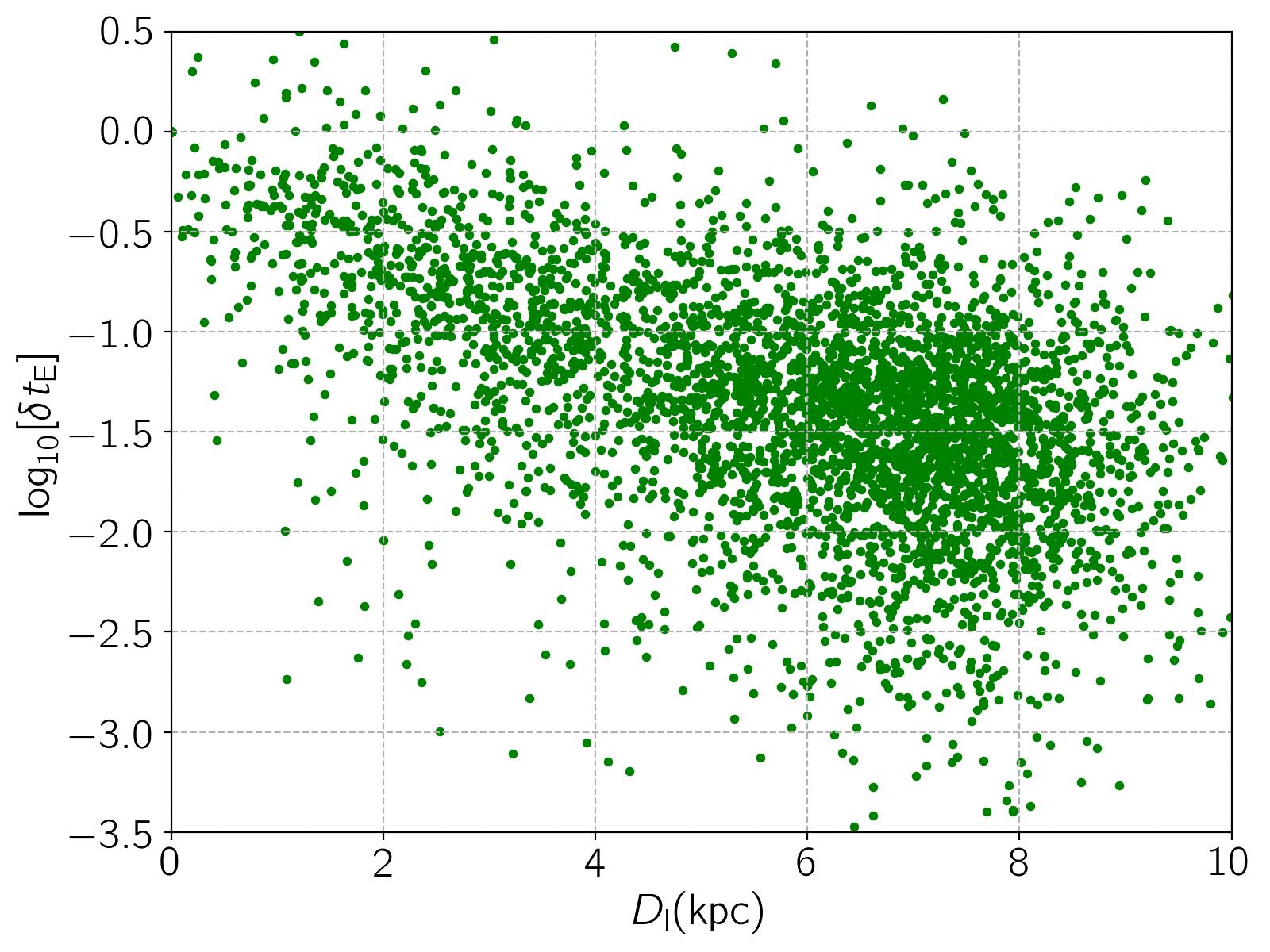}
\includegraphics[width=0.4\textwidth]{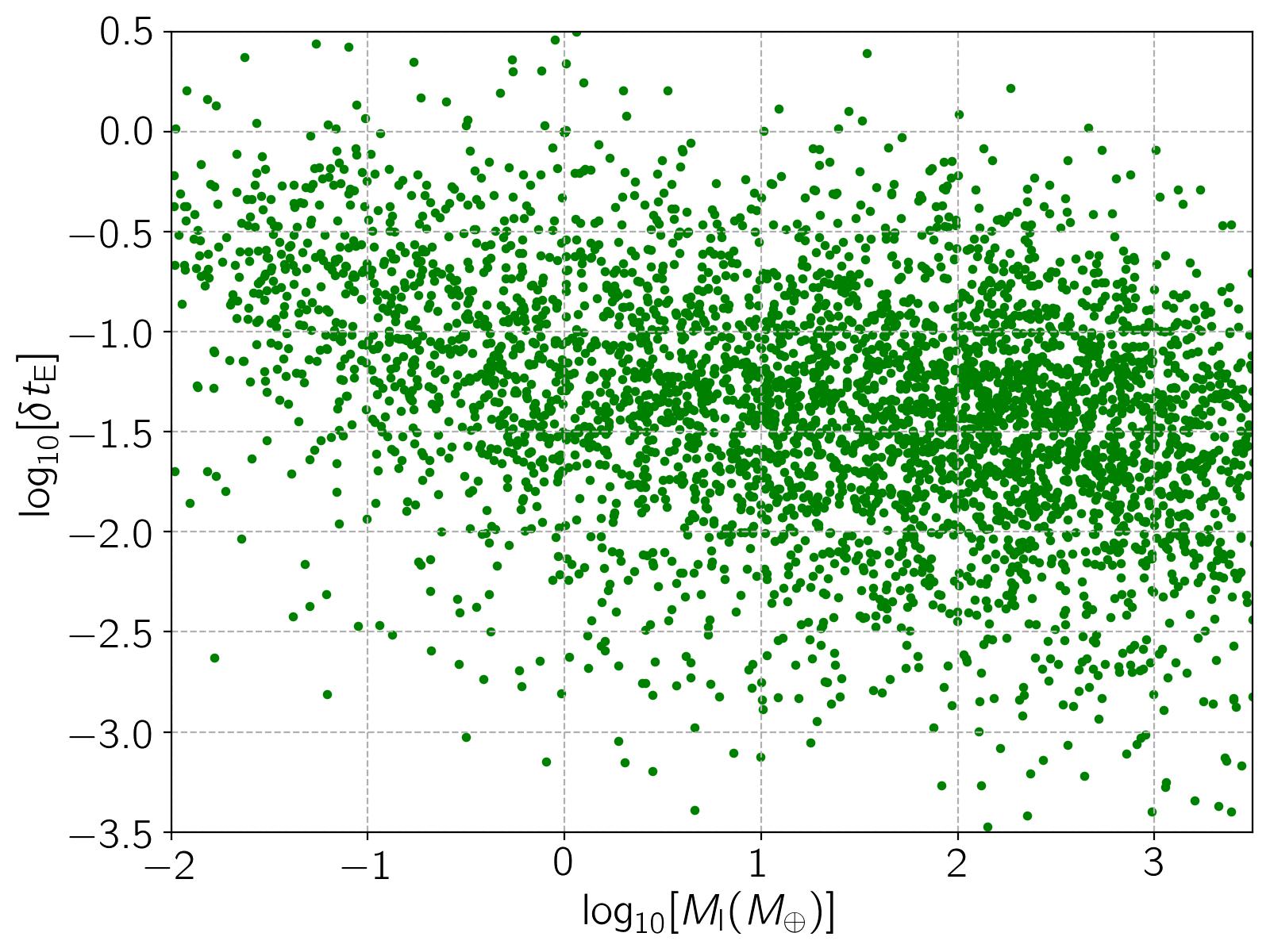}
\caption{The scatter plot of the relative deviation in the Einstein crossing time $\delta t_{\rm E}$ versus two parameters $D_{\rm l}$, and $\log_{10}[M_{\rm l}(M_{\oplus})]$.}
\label{fig6}
\end{figure}

\section{Statistical results}\label{sec3}
After indicating the best-fitted simple models for the simulated events with $\Delta \chi^{2}>100$, we statistically investigate the fractions of events in which the best-fitted lensing parameters are considerably different from the real ones. These fractions are reported in Table \ref{tab2}. In this regard, we define the relative variations in the lensing parameters, e.g., $\delta t_{\rm E}=|t_{\rm E,~r}-t_{\rm E,~b}|/t_{\rm E, r}$. Here, $t_{\rm E,~r}$, and $t_{\rm E,~b}$ are the Einstein crossing times from the real lensing model with parallax and the best-fitted simple model, respectively. Hence, the indexes $r$ and $b$ (throughout the paper) are referred to the real and best-fitted models. In the same way, we define the relative deviations in other lensing parameters as: $\delta \rho_{\star}=|\rho_{\star,~r}-\rho_{\star,~b}| /\rho_{\star, r}$,  $\delta u_{0}=|u_{0,~r}-u_{0,~b}| /u_{0, r}$, $\delta t_{0}=|t_{0,~r}-t_{0,~b}| /t_{0, r}$, and $\delta f_{\rm bl}=|f_{\rm bl,~r}-f_{\rm bl,~b}| /f_{\rm bl, r}$. 
 
According to Table \ref{tab2}, the fraction of events in which the times of closest approach are shifted by $\delta t_{0}\gtrsim 0.3$ is only $0.2\%$. The most-affected parameter (by parallax) is $\rho_{\star}$. In $56.9\%$ of events with $\Delta \chi^{2}>100$ the relative deviation $\delta \rho_{\star}$ is larger than $0.3$. $8.0\%$ of the parallax-affected events have $\delta t_{\rm E}>0.3$. Deflections in $t_{\rm E}$, and $\rho_{\star}$ lead to misinterpreting the lens object (its mass and distance). According to the simulation, we found that in $4.9\%$ of simulated parallax-affected events with $\Delta \chi^{2}>100$, both relative deviations $\delta t_{\rm E}$ and $\delta \rho_{\star}$ were greater than $0.3$. 

The fraction of parallax-affected events as a function of four relevant parameters are plotted in Figure \ref{fig5}. In each panel, three depicted lines are corresponding to three different criteria to extract the affected events which are: (i) gray solid lines show the fractions of events with $\Delta \chi^{2}>100$, (ii) red dashed lines represent the fraction of events with $\Delta \chi^{2}>100$ and $\delta t_{\rm E}>0.02$, and (iii) green dot-dashed lines reveal the fraction of events with $\Delta \chi^{2}>100$ and $\delta \rho_{\star}>0.02$. Four points from different panels of this figure are mentioned in the following.  

\begin{itemize}[leftmargin=2.0mm]
\item According to the top-left panel: the events with higher $\pi_{\rm{rel}}$ values (due to closer lens objects) are more affected by the parallax effect. According to the discussion in the previous sections, in short-duration microlensing events due to FFPs the parallax-induced deviation $\boldsymbol{\delta u}$ is almost a straight line, and $\delta u \propto \pi_{\rm{rel}}$. By increasing $\pi_{\rm{rel}}$, the relative deviations in both $t_{\rm E}$, and $\rho_{\star}$ enhance. We also represent the scatter plot of $\delta t_{\rm E}$ versus $D_{\rm l}$ in the top panel of Figure \ref{fig6} which emphasizes this point as well.  

\item More massive lens objects produce longer microlensing events with lower finite-source size and lower $\pi_{\rm E}$. Hence, the lens mass has three effects on parallax-induced deviations. (i) Generally, longer events have higher numbers of data points, and $\Delta \chi^{2}$ values. (ii) For longer microlensing events the approximations in Equations \ref{twot} are not valid, and by considering more terms in these expansions, $\delta u$ will reduce by the lens mass. (iii) The events with lower $\rho_{\star}$ values have on average higher magnification factors and are more sensitive to variations in parameters, specially $\rho_{\star}$. Considering all of these points, longer events due to more massive lens objects have lower $\delta t_{\rm E}$ (see also the bottom panel of Figure \ref{fig6}), but higher $\delta \rho_{\star}$, as depicted in the top-right panel of Figure \ref{fig5}. 

\item According the bottom-left panel of Figure \ref{fig5}, the events with larger finite-source sizes (with smaller $\theta_{\rm E}$ values) are less affected by parallax. As discussed in the previous section, magnification factor due to large $\rho_{\star}$ is low and depends approximately on the source size itself \citep{1996Gouldfinite,2003ApJAgol}, and different lens trajectories passing from large source disks have almost similar light curves. On the other hand, the magnification factor for such events is not high (very close to one) which causes on average higher photometric errors, and lower $\Delta \chi^{2}$ values. We note that considering (i) $\theta_{\rm E}=\sqrt{\kappa~M_{\rm l}~\pi_{\rm{rel}}}$, and (ii) two factors of the lens mass and the relative parallax have inverse effects on $\delta t_{\rm E}$, hence $\delta t_{\rm E}$ is not much sensitive to $\theta_{\rm E}$.  

\item The last panel of Figure \ref{fig5} emphasizes that the relative deviation $\delta t_{\rm E}$ additionally depends on the angle between the lens-source relative trajectory and the parallax vector ($|\xi-\phi_{0}|$). When $\boldsymbol{u}_{\odot}$ and $\boldsymbol{\delta u}$ are parallel ($| \xi-\phi_{0}|\simeq 0, 180$) the width of light curves and specially $t_{\rm E}$ changes considerably. 
\end{itemize}

\begin{figure*}
	\centering	
	\includegraphics[width=0.9\textwidth]{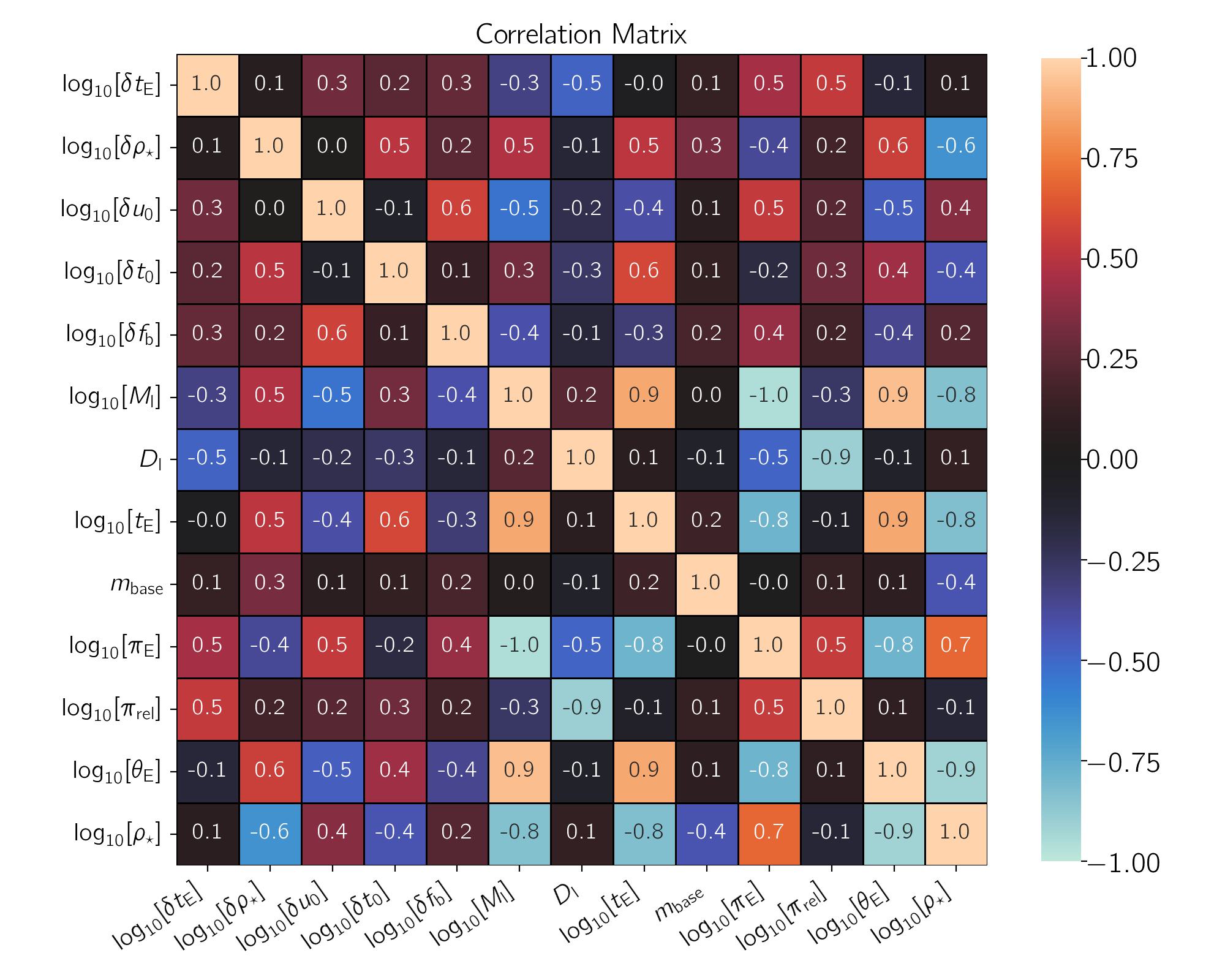}
	\caption{The correlation matrix between the relative deviations in the lensing parameters (the first five rows) due to the indiscernible parallax effects and the relevant and physical parameters of simulated events(the other rows).}
	\label{fig1t}
\end{figure*}

There are some other correlations between the real parameters and the relative deviations in the lensing parameters. These correlations reveal in what kind of short-duration microlensing events the indiscernible parallax effect rather changes each of lensing parameters. In Figure \ref{fig1t} we show the correlation matrix between relative deviations in the lensing parameters in the logarithmic scale (i.e., $\log_{10}[\delta t_{\rm E}]$, $\log_{10}[\delta \rho_{\star}]$, $\log_{10}[\delta u_{0}]$, $\log_{10}[\delta t_{0}]$, and $\log_{10}[f_{\rm bl}]$) and the eight relevant and physical parameters of simulated events (include $\log_{10}[M_{\rm l}]$, $D_{\rm l}$, $\log_{10}[t_{\rm E}]$, $m_{\rm{base}}$, $\log_{10}[\pi_{\rm E}]$, $\log_{10}[\pi_{\rm{rel}}]$, $\log_{10}[\theta_{\rm E}]$, and $\log_{10}[\rho_{\star}]$). 
The correlation coefficients which are mentioned in the color bar are in the range $[-1,~ +1]$. The positive and negative correlations between two entrances mean values of the first one are rising as those in the second one increase, and vice versa. The zero correlation coefficient means there is no correlation between the given entrances.   

\noindent The highest correlation for the relative deviations for the Einstein crossing time is $0.5$ and with $D_{\rm l}$, and as a result $\log_{10}[\pi_{\rm rel}]$ and $\log_{10}[\pi_{\rm E}]$, as emphasized in the previous section.    

\noindent The relative deviation in the source radius is highly correlated with $\log_{10}[\theta_{\rm E}]$ and $\log_{10}[\rho_{\star}]$ with the correlation coefficient $\pm 0.6$. There are weaker and positive correlations between it and $\log_{10}[t_{\rm E}]$, and the lens mass.

\noindent The relative deviation in the lens impact parameter ($\delta u_{0}$) is correlated with $\log_{10}[\pi_{\rm E}]$ and $\log_{10}[M_{\rm l}]$ with the correlation coefficients $\pm 0.5$. 

\noindent In the regard of relative deviation in the time of closest approach, the highest correlation is $0.6$ and with $\log_{10}[t_{\rm E}]$. 
Finally, the relative deviation in the blending factor is highly correlated with $\delta u_{0}$ which is $0.6$. We note that the blending effect makes a degeneracy between parameters in microlensing events due to FFPs specially the ones with considerable finite-source sizes \citep{2020AJMroz,2023sajadiannumerical}, which makes these correlations between relative deviations.
The microlensing events due to fainter source stars have higher relative deviations in $t_{\rm E}$, $\rho_{\star}$, $t_{0}$, and $f_{\rm b}$. 

To estimate the number of FFPs will be detected by the \wfirst\ telescope through its microlensing survey so that their parameters are misinterpreted because of the indiscernible parallax effect, we use the predictions offered by \citet{2020AJjohnson}. Accordingly, the \wfirst\ telescope will detect $N_{\rm{FFP}}=897$ FFPs in the mass range $[0.1 M_{\oplus},~1000 M_{\oplus}]$ by assuming a log-uniform mass function  for them (see Table (2) of \citet{2020AJjohnson}). According to our simulation in the same mass range, $f_{1}= 36.5\%$ of such events have $\Delta \chi^{2}>100$. In $f_{2}=13.9,~3.1,~1.1\%$ of these events, indistinguishable parallax effects make the relative deviations in both $t_{\rm E}$ and $\rho_{\star}$ larger than $0.1$,~$0.3$, and $0.5$ respectively. The number of parallax-affected FFPs detectable by the \wfirst\ telescope can be estimated by $N= N_{\rm{FFP}} \times f_{1} \times f_{2}$. Hence, 46, 10, 4 of these FFPs which are discovered through \wfirst\ microlensing survey are affected by invisible parallax so that both $\delta t_{\rm E}$, and $\delta \rho_{\star}$ are larger than 0.1, 0.3, 0.5, respectively.

\section{Conclusions}\label{sec4}
Our Galaxy contains trillions of FFPs\citep{2023AJ108sumi}. Gravitational microlensing is the powerful technique to discover such isolated and dark objects even low-mass ones \citep[see, e.g., ][]{Sumi2011Natur,Mroz2017Natur}. Although, the James Web Space Telescope (JWST) can find planetary-mass objects as small as $0.6$ Jupiter-mass which are very young and located in active stellar formation regions \citep{2023pearson}, gravitational microlensing is still the only way to find cold and dark small bodies in our Galaxy regardless to their ages and locations.

In the short-duration microlensing events due to FFPs with planetary masses (i.e., $M_{\rm l} \in [0.01 M_{\oplus}, 15 M_{\rm J}]$ where, $M_{\rm J}$ is the Jupiter's mass), the parallax amplitudes are higher than those in common events (due to M-dwarfs with the mass $\sim 0.3 M_{\odot}$) by $[5,~3000]$ times. On the other hand, the duration of such events is very short as it is scaled by $t_{\rm E} \propto \sqrt{M_{\rm l}}$, mostly up to a few days. Considering the orbital period of the Earth rotation around the Sun ($P_{\oplus}$), in such short-duration events discerning parallax effect is barely possible.     

\noindent In the first glance, the parallax effect is not recognizable in such events and can not help resolving the microlensing degeneracy. In another glance, although the duration of these events are short and we can not model and discern the annual parallax, this effect changes the shapes of observed light curves and their lensing parameters, because of large $\pi_{\rm E}$ values in these events. 

In a short-duration microlensing events due to FFPs, if $t_{\rm E}\ll P_{\oplus}$ we can expand the parallax-induced deviation in the lens-source relative trajectory ($\boldsymbol{\delta u}$). By considering only the first-order term in $t_{\rm E}/P_{\oplus}$, this deviation makes a straight line without bending whose size is $\propto \pi_{\rm{rel}}$. Hence, the observer motion around the Sun alters the lens-source relative trajectory in this event as $\boldsymbol{u}_{\rm o}=\boldsymbol{u}_{\odot}+\boldsymbol{\delta u}$, and as a results its light curve. The point is that both light curves $A(u_{\rm o})$ and $A(u_{\odot})$ are simple with different lensing parameters. If only one observer takes data from these short-duration events due to FFPs, the indiscernible parallax effect deviates the best-fitted lensing parameters from their real values. 

In this work, we evaluated the parallax-induced deviations in these events statistically. We simulated microlensing events due to FFPs which have $t_{\rm E}<10$ days and assumed that the \wfirst\ telescope would be the only observer for these events. We expect that some of short-duration microlensing events which will be alerted by \wfirst\ are not followed up by other telescopes (rotating the Sun from different orbits), for instance the events with very short time scales, or follow up telescopes may have other priorities at the time of the \wfirst\ observations. 

$46.3\%$ of these simulated events had $\Delta \chi^{2}=| \chi^{2}_{\rm{real}}-\chi^{2}_{\rm{without}}|>100$ (where $\chi^{2}_{\rm{real}}$ and $\chi^{2}_{\rm{without}}$ are the $\chi^{2}$ values from fitting the real model with and without the parallax effect). These events mostly were due to FFPs close to the observer $D_{\rm l}\lesssim 3$ kpc (which had $\pi_{\rm{rel}}\gtrsim 0.2$ mas). Also the events due to either faint or highly blended source stars, or ones with $\rho_{\star}\gtrsim 3$ are less affected by parallax (see Figure \ref{fig2}). We found that when the parallax vector and the lens-source trajectory were parallel, the widths of light curves changed significantly (see Figure \ref{fig3}).

For parallax-affected events with $\delta \chi^{2}>100$, we inferred the best-fitted simple Paczy\'nski microlensing models and evaluated the relative deviations in the lensing parameters, i.e., $\delta t_{\rm E}$, $\delta \rho_{\star}$, $\delta u_{0}$, $\delta t_{0}$, and $\delta f_{\rm bl}$. We concluded that $\rho_{\star}$ is the most-affected parameter so that $\delta \rho_{\star}$ was $>0.3$ in $56.9\%$ of events. Also, the time of the closest approach $t_{0}$ was the least-affected parameter, so that $\delta t_{0}>0.3$ occurred only in $0.2\%$ events.

We estimated that 46 of microlensing events due to FFPs which will be discovered by \wfirst\ are affected by missing parallax so that both relative deviations in $t_{\rm E}$ and $\rho_{\star}$ are larger than $0.1$. This number of FFPs whose light curves (and their lensing parameters) are affected by the parallax effect reveals the importance of doing simultaneous and dense (with a short cadence) observations from microlensing events alerted by \wfirst\ to capture the parallax deviations in short-duration microlensing events due to FFPs. In this regard, the extra observations should be done with the observers rotating the Sun in different orbits from the \wfirst\ orbit.

\small
All simulations that have been done for this paper are available in the GitHub and Zenodo addresses: \url{https://github.com/SSajadian54/FFPs_parallax/}, and and \url{https://zenodo.org/record/8342045}\citep{sajadian_2024_10827255}.


\bibliographystyle{aasjournal}
\bibliography{paper}{}
\end{document}